\documentclass[acmtog,authorversion=true]{acmart}

\usepackage{booktabs} 
\usepackage[final]{pdfpages}

\usepackage{environ}
\usepackage{natbib}
\setcitestyle{square}
\usepackage[ruled]{algorithm2e} 
\usepackage[f]{esvect}
\usepackage{amsfonts,amssymb,calc,amsmath}
\usepackage[usenames,dvipsnames]{pstricks}
\usepackage{graphicx}
\usepackage{caption}
\usepackage{subcaption}
\usepackage{xspace}
\usepackage{cancel}
\usepackage{units}
\usepackage[USenglish]{babel}
\usepackage{mathtools}
\usepackage{booktabs}
\usepackage{textcomp}
\usepackage{xr}
\usepackage{verbatim}
\usepackage{array}

\usepackage{footnote}
\makesavenoteenv{tabular}
\makesavenoteenv{table}

\usepackage[many]{tcolorbox}
\usepackage{empheq}

\graphicspath{ {./figures/} }

\SetAlFnt{\small}
\SetAlCapFnt{\small}
\SetAlCapNameFnt{\small}
\SetAlCapHSkip{0pt}
\IncMargin{-\parindent}

\acmJournal{TOG}
\acmVolume{37}
\acmNumber{2}
\acmArticle{20}
\acmYear{2018}
\acmMonth{2}

\setcopyright{none}

\acmDOI{10.1145/3185225}

\usepackage{siunitx}

\definecolor{red}{rgb}{0.8,0,0}
\definecolor{black}{rgb}{0,0,0}
\definecolor{purered}{rgb}{1,0,0}
\definecolor{darkred}{rgb}{0.6,0,0}
\definecolor{green}{rgb}{0.0,0.5,0}
\definecolor{blue}{rgb}{0,0,0.75}
\definecolor{darkblue}{rgb}{0,0,0.55}
\definecolor{orange}{rgb}{0.9,0.3,0.1}
\definecolor{purple}{rgb}{0.8,0.0,0.8}
\definecolor{cyan}{rgb}{0.0,0.7,0.7}

\newcommand{\mat}[1]{{\begin{pmatrix} #1 \end{pmatrix}}}

\newcommand{\diegoc}[1]{}
\newcommand{\julioc}[1]{}
\newcommand{\juliotxt}[1]{}
\newcommand{\julio}[1]{}
\newcommand{\recheck}[1]{}
\newcommand{\missing}[1]{}
\newcommand{\wojciech}[1]{}
\newcommand{\adrianc}[1]{}
\newcommand{\major}[1]{{\leavevmode \color{black}{#1}}}
\newcommand{\minor}[1]{{\leavevmode \color{black}{#1}}}
\newcommand{\remove}[1]{}

\def\equationautorefname~#1\null{%
  Equation~(#1)\null
}

\renewcommand{\leftarrow}{\textcolor{red}{***}}
\newcommand{\Fig}[1]{\autoref{fig:#1}}
\newcommand{\Figs}[2]{Figures~\ref{fig:#1} and \ref{fig:#2}}
\newcommand{\Tab}[1]{\autoref{tab:#1}}
\newcommand{\Eq}[1]{\autoref{eq:#1}}
\newcommand{\Eqs}[2]{Equations~\eqref{eq:#1} and \eqref{eq:#2}}

\newcommand{\Sec}[1]{\autoref{sec:#1}}

\newcommand{\Apx}[1]{\autoref{ap:#1}}
\newcommand{\Apxx}[2]{Appendices~\ref{ap:#1} and \ref{ap:#2}}

\newcommand{\EqsRange}[2]{Equations~(\ref{eq:#1}--\ref{eq:#2})}

\newcommand{\balign}[1]{\begin{tcolorbox} [colback=blue!5!white,size=small,width=\linewidth, boxsep=2pt, bottom=1pt, top=-3pt]\begin{align} #1 \end{align}\end{tcolorbox}}
\newcommand{\fr}{\frac}

\newcommand{\larrow}{\leftarrow}

\newcommand{\pdf}{\textrm{pdf}}

	\newcommand{\DD}{{\mathrm{2D}}}
\newcommand{\DDD}{{\mathrm{3D}}}
\newcommand{\abs}{\mu_a} 
\newcommand{\sca}{\mu_s} 
\newcommand{\ext}{\mu_t} 

\newcommand{\pf}{f} 
\newcommand{\pfs}{f_s} 
\newcommand{\crossSection}{\sigma} 

\newcommand{\omegaout}{\ensuremath{\vec{\omega}_o}\xspace}
\newcommand{\omegain}{\ensuremath{\vec{\omega}_i}\xspace}
\newcommand{\wout}{\omegaout}
\newcommand{\win}{\omegain}
\newcommand{\Tr}{{T_r}}
\newcommand{\eqbreak}{\nonumber \\}
\newcommand{\x}{\ensuremath{\mathbf{x}}\xspace}
\newcommand{\y}{\ensuremath{\mathbf{y}}\xspace}

\newcommand{\rvec}{\vec{\ensuremath{\mathbf{r}}}\xspace}
\newcommand{\rlen}{\ensuremath{\mathrm{r}}\xspace}
\newcommand{\xpr}{\ensuremath{\mathbf{x}^\prime}\xspace}
\newcommand{\xm}{\ensuremath{\x}\xspace}
\newcommand{\ym}{\ensuremath{\y}\xspace}

\newcommand{\VRegion}{\ensuremath{\aleph}\xspace}
\newcommand{\xs}{\ensuremath{\x_{s}}\xspace}
\newcommand{\ys}{\ensuremath{\y_{\! s}}\xspace}

\newcommand{\yw}{\ensuremath{\y}\xspace}

\newcommand{\Deltaxpr}{\Delta_{\xpr}}
\newcommand{\DeltaxprT}{\Delta_{\xpr}^\intercal}

\newcommand{\sDistance}{\ring}

\newcommand{\xy}{\vv{\x\y}}
\newcommand{\xyT}{\xy^\intercal}
\newcommand{\xyi}{\vv{\x\y_i}}

\newcommand{\yx}{\vv{\y\x}}
\newcommand{\yxT}{\yx^\intercal }

\newcommand{\ny}{\vec{\mathbf{n}}_\y}

\newcommand{\nx}{\vec{\mathbf{n}}_\x}

\newcommand{\vvec}{\vec{\mathbf{v}}}

\newcommand{\normyx}{\|\yx\|}

\newcommand{\cosy}{\cos{\theta_\y}}
\newcommand{\thetap}{\theta^\prime}

\newcommand{\costhetap}{\cos{\thetap}}
\newcommand{\Geom}{G}

\newcommand{\GeomMs}{\Geom}

\newcommand{\mytransp}{\intercal}

\renewcommand{\th}{\ensuremath{^{\text{th}}}}
\newcommand{\GR}{\nabla}

\newcommand{\GRx}{\GR}

\newcommand{\GRxT}{\GR^\mytransp}
\newcommand{\HS}{\mathbf{H}}

\newcommand{\HSx}{\HS}

\newcommand{\JCx}{\mathbf{J}}
\newcommand{\Segments}{\mathcal{L}}
\newcommand{\seg}{\ell}
\newcommand{\segj}{{\seg_j}}

\newcommand{\Rings}{\mathcal{R}}
\newcommand{\ring}{r}
\newcommand{\sErrorTolerance}{\varepsilon}
\newcommand{\sError}{\epsilon}
\newcommand{\estErr}{\hat{\sError}}
\newcommand{\estRelError}{\estErr'}

\newcommand{\Rad}{\ensuremath{L}\xspace}
\newcommand{\Rado}{\ensuremath{L}\xspace}
\newcommand{\RadoIdx}[1]{\ensuremath{L_{#1}}\xspace}
\newcommand{\Radi}{\ensuremath{L_i}\xspace}

\newcommand{\estDD}{(\xm, \seg)} 
\newcommand{\estTrDD}{{\Tr}\estDD}

\newcommand{\estRadDD}{\Radi(\xm \larrow \seg)}

\newcommand{\redRad}{\Rad_r}
\newcommand{\omegal}{\ensuremath{\vec{\omega}_\seg}}
\newcommand{\omegalj}{\ensuremath{\vec{\omega}_{\segj}}}
\newcommand{\xl}{\ensuremath{\y_\seg}}
\newcommand{\xlj}{\ensuremath{\y_\segj}}

\newcommand{\Vis}{V}
\newcommand{\Irr}{\ensuremath{E}\xspace}
\newcommand{\IrrDD}{\ensuremath{\Irr_{\DD}}\xspace}

\newcommand{\FFDDM}{F_{\seg}}

\newcommand{\FFDDMj}{F_{\segj}}
\newcommand{\FFDDDM}{F_{\tri}}

\newcommand{\tri}{\triangle}

\newcommand{\diff}{\mathrm{d}}
\newcommand{\diffseg}{\diff \seg}

\newcommand{\norm}[1]{\| #1 \|}
\newcommand{\SATriangle}{\Omega}
\newcommand{\absvalue}[1]{\lvert #1 \rvert}
\newcommand{\dotp}[2]{#1\! \cdot\! #2}
\newcommand{\dotpp}[2]{\left(\dotp{#1}{#2}\right)}
\newcommand{\crossp}[2]{#1 \times #2}
\newcommand{\crosspp}[2]{\left(\crossp{#1}{#2}\right)}

\begin{document}
	
\title{Second-Order Occlusion-Aware Volumetric Radiance Caching}

\author{Julio Marco}
\affiliation{%
  \institution{Universidad de Zaragoza, I3A}
}
\author{Adrian Jarabo}
\affiliation{%
  \institution{Universidad de Zaragoza, I3A}
}

\author{Wojciech Jarosz}
\affiliation{%
  \institution{Dartmouth College}
}

\author{Diego Gutierrez}
\affiliation{%
  \institution{Universidad de Zaragoza, I3A}
}

\renewcommand\shortauthors{Marco et al.}

\begin{abstract}
%
\major{We present a second-order gradient analysis of light transport in participating media and use this to develop an improved radiance caching algorithm for volumetric light transport.} We adaptively sample and interpolate radiance from sparse points in the medium using a second-order Hessian-based error metric to determine when interpolation is appropriate. We derive our metric from each point's incoming light field, computed by using a proxy triangulation-based representation of the radiance reflected by the surrounding medium and geometry. We use this representation to efficiently compute the first- and second-order derivatives of the radiance at the cache points while accounting for occlusion changes.
\major{We also propose a self-contained two-dimensional model for light transport in media and use it to validate and analyze our approach, demonstrating that our method outperforms previous radiance caching algorithms both in terms of accurate derivative estimates and final radiance extrapolation. We generalize these findings to practical three-dimensional scenarios, where we show improved results while reducing computation time by up to 30\% compared to previous work.}


\end{abstract}

%
%

\ccsdesc[500]{Computing methodologies~Ray tracing}

%
%

\keywords{global illumination, rendering, irradiance caching, participating media, radiance derivatives}
\thanks{}
\maketitle

\begin{figure*}[t]
\begin{subfigure}[t]{.6\textwidth}
  \centering
  \includegraphics[width=\textwidth]{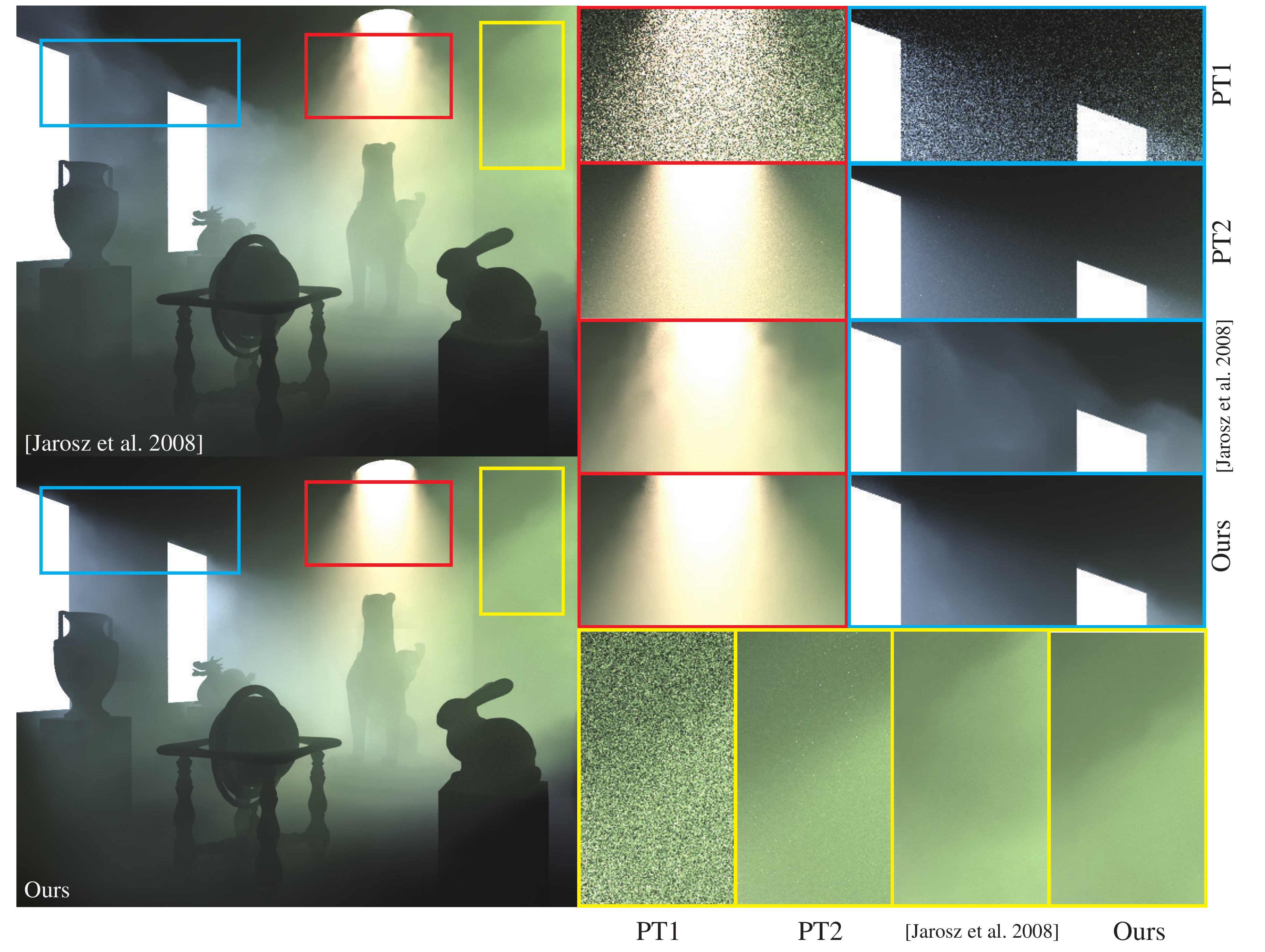}
\caption{}
\label{fig:statuesRenders}
\end{subfigure}
\hfill
\begin{subfigure}[t]{.39\textwidth}
  \centering
  \includegraphics[width=\textwidth]{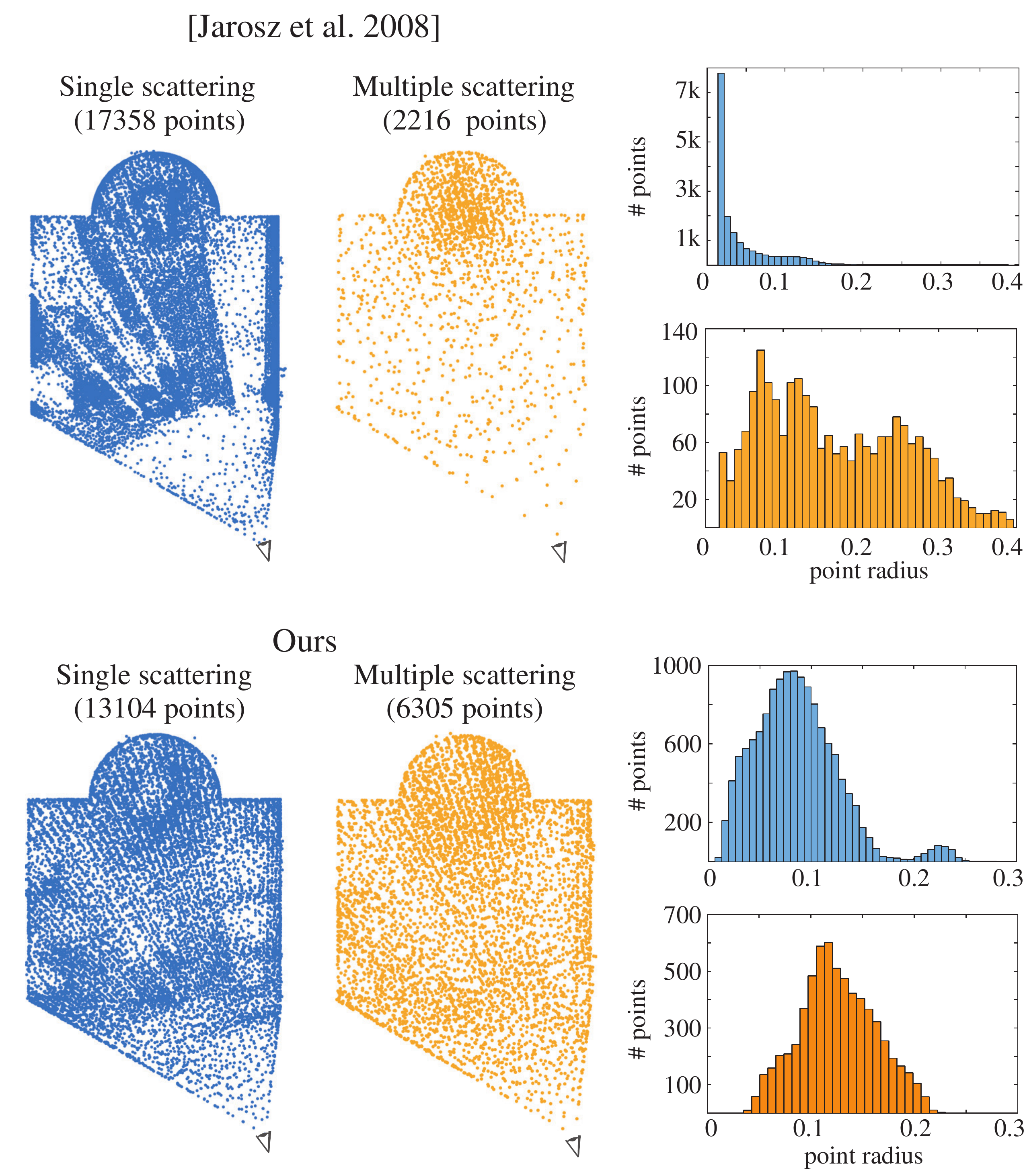}
\caption{}
\label{fig:statuesHistograms}
\end{subfigure}

\caption{
%
\textit{Statues} scene rendered with both single and multiple scattering. Radiance at surfaces is excluded for illustration purposes (please refer to the digital version for accurate visualization).
%
(a) \textbf{PT1:} Path tracing, 2k samples/pixel, 2h. \textbf{PT2:} Path tracing, 500k samples/pixel, 500h. \textbf{\protect\cite{jarosz08radiance}:} Occlusion-unaware, gradient-based error metric, $\sim$19k cache points, 16k samples/cache, 155 minutes. \textbf{Ours:} Occlusion-aware, Hessian-based metric, $\sim$19k cache points, 16k samples/cache, 154 minutes.
(b) Cached point distribution as seen from above for both single  and multiple scattering.
Ignoring visibility derivatives fails at representing high-frequency shadows from the windows (a, blue and yellow) due to poor cache distribution, as well as other rapid radiance changes (a, red) in areas with good cache distribution, due to imprecise extrapolation during reconstruction.
In contrast, our occlusion-aware Hessian-based method correctly handles these higher-frequency features by improving the sample distribution, as well as the reconstruction.
The occlusion-unaware approach (b, top-left) concentrates the samples excessively near the surfaces, usually reaching the cache minimum radius (see top-right histograms), but ignoring occlusion changes throughout the scene. Using occlusion-aware first- and second-order derivatives, our method predicts the error introduced by gradient extrapolation more robustly, increasing cache density in regions where gradients change rapidly (b, bottom). \label{fig:statues}
}
\end{figure*}
\vspace{-0.5em}
\section{Introduction}

Accurately simulating the complex lighting effects produced by participating media in the presence of arbitrary geometry remains a challenging task.
Monte Carlo-based methods like path tracing numerically approximate the radiative transfer equation (RTE)~\cite{Chandrasekhar1960} by stochastically sampling radiance in the medium. These approaches can handle complex geometry and general scattering properties, but since they lack memory and are largely blind to the radiance signal, they perform many redundant computations leading to high cost.
A common strategy to increase efficiency is to adaptively sample radiance based on its frequency content, limiting the sampling density in regions where radiance barely changes, and placing more samples in regions with higher frequency variation~\citep{zwicker15star}.

Based on this principle, \emph{volumetric radiance caching}~\citep{jarosz08radiance} computes and stores radiance at sparse cache points in the medium, and uses these samples to reconstruct radiance at nearby locations whenever possible. The method is based on first-order translational derivatives of the radiance, which are used to i) determine how far away a cache point can be reused while controlling error, and ii) improve reconstruction quality by extrapolating the cached radiance values along their gradients.
Unfortunately, since the gradient derivations ignore occlusion/visibility changes, the method fails in scenes containing occluders where changes in visibility are the dominant factor in local radiance behavior. Moreover, the reconstruction and error metric both rely on the same gradient estimates and ignore variations caused by higher-order derivatives. These factors lead to suboptimal cache point distributions, which fail to properly sample high-frequency features such as occlusions, while simultaneously oversampling other regions of the scene. This results in reduced efficiency and visible rendering artifacts.

Second-order illumination derivatives have proven to be a powerful and principled tool for sparsely sampling and interpolating surface irradiance~\citep{jarosz12theory,schwarzhaupt12practical}, as well as controlling error in density estimation techniques~\citep{hachisuka10progressive,Kaplanyan2013APPM,Belcour2014}. Inspired by these recent developments, we propose a new second-order, \textit{occlusion-aware} radiance caching method for participating media which overcomes the limitations of current state-of-the-art methods.

To this end, we introduce a novel approach to compute first- and second-order occlusion-aware derivatives of both single and multiple scattering, and \major{generalize the Hessian-based metric of \citet{schwarzhaupt12practical} for controlling the error introduced by first-order extrapolation of media radiance.}
In addition, we extend recent work on 2D radiometry, currently limited to surfaces~\citep{jarosz12theory}, and derive a 2D theory of light transport in participating media. We use this framework to illustrate and analyze the limitations of the state of the art, as well as the benefits of our proposed method.
%
We demonstrate the generality of our approach by deriving occlusion-aware derivatives of 3D media radiance and applying our Hessian-based metric to 3D cache distributions, showing that the benefits predicted by our 2D analysis hold equally in 3D.
%
Our approach improves volumetric cache point distributions \major{in isotropic homogeneous media}, providing a significantly more accurate reconstruction of difficult high-frequency features, as \autoref{fig:statues} shows.

\vspace{-1em}
\section{Related Work}
We summarize here existing work on radiance caching methods as well as other techniques that leverage illumination derivatives to improve Monte Carlo rendering.
For a general overview of scattering and existing adaptive sampling and reconstruction techniques, we refer the reader to other recent sources of information~\cite{Gutierrez2008course,zwicker15star}.

\textbf{Radiance caching: }
Irradiance caching was originally proposed by \citet{Ward1988} to accelerate indirect illumination in Lambertian scenes. The method computes and caches indirect irradiance only at a sparse set of points in the scene, and extrapolates or interpolates these values whenever possible from cache points deemed to be sufficiently close by. Since indirect illumination changes slowly across Lambertian surfaces, the costly irradiance calculation can often be reused over large parts of the image, substantially accelerating rendering. There has been a wealth of improvements to irradiance caching, but we discuss only the most relevant follow-up work and refer to \citet{Krivanek:2009:PGI} for a more complete survey.

\Citet{wardheckbert1992} significantly improved reconstruction by leveraging gradient information, and \citet{Krivanek2006clamping} incorporated heuristics to improve error estimation (and therefore quality) during adaptive caching. K{\v{r}}iv{\'{a}}nek and colleagues~\citeyear{Krivanek2005caching,Krivanek2005improvedcaching} also extended irradiance caching to handle moderately glossy, non-Lambertian surfaces. \major{\citet{Herzog2009} used anisotropic cache points based on the orientation of the illumination gradient.} All these \major{methods only considered \emph{surface} light transport}.

\Citet{jarosz08radiance} proposed \textit{volumetric} radiance caching, which accelerates single and multiple scattering in participating media. They proposed an error metric based on the first-order derivative of the radiance, but their formulation ignored volumetric occlusion changes. In follow-up work, \citet{jarosz08irradiance} derived occlusion-aware gradients, but only of surface illumination in the presence of absorbing and scattering media, ignoring gradients of the media radiance itself. Both approaches are prone to suboptimal cache point distributions and visible artifacts since they ignore higher order derivatives or occlusion changes in media. Our work addresses both of these issues. \major{\citet{Ribardiere2011} proposed using anisotropic cache points and a second-order expansion for radiance reconstruction. Their approach, however, did not consider visibility changes due to their point-to-point computation of derivatives.}

Recently, \citet{jarosz12theory} and follow-up work~\citep{schwarzhaupt12practical} made significant progress in  heuristics-free error control for surface irradiance caching by formulating error in terms of second-order derivatives. In particular, \citet{schwarzhaupt12practical} proposed a novel radiometrically equivalent formulation of irradiance gradients and Hessians, which properly accounted for occlusions. The authors used these for extrapolation and principled error control, respectively. We extend these ideas and apply them to light transport in participating media, deriving first- and second-order occlusion-aware derivatives for improved reconstruction and principled error control in volumetric radiance caching.

%
\begin{table*}[t!]
\caption{\label{tab:notation}
Notation for the optical properties of participating media, and their differences in 3D and 2D.} 
\resizebox{\textwidth}{!}{%
\begin{tabular}{lclclc}
\toprule
&  & \multicolumn{2}{c}{3D} &\multicolumn{2}{c}{2D}\\
\cmidrule(r){3-4} \cmidrule(r){5-6}
Quantity & Symbol & Expression & Units & Expression & Units\\
\midrule
 Particle density &  $\rho$  & Particles per unit volume & \si{\per\cubic\meter} & Particles per unit area & \unit{\si{\per\square\meter}} \\
Cross-section &  $\crossSection$  & Area & \unit{\si{\square\meter}} & Length & \unit{\si{\meter}} \\
Scattering coefficient & $\sca$ & Probability density  per differential length & \si{\per\meter} & Probability density per differential length & \si{\per\meter} \\
Absorption coefficient & $\abs$ & Probability density per differential length & \si{\per\meter} & Probability density per differential length & \si{\per\meter} \\
Extinction coefficient & $\ext$ & $\ext = \abs + \sca$  & \si{\per\meter} & $\ext = \abs + \sca$  & \si{\per\meter} \\
Transmittance & $T_r$ & $T_r(\x_1, \x_2)  = \exp({-\int_{\x_1}^{\x_2}\!{\ext(\x)}\diff\x})$  & unitless & $T_r(\x_1, \x_2)  = \exp({-\int_{\x_1}^{\x_2}\!{\ext(\x)}\diff\x})$  & unitless \\
Phase Function & ${\pf(\x, \omegain,\omegaout)}$ & Angular scattering of light at a point & \si{\per\steradian} & Angular scattering of light at a point &  \si{\per\radian} \\
\bottomrule
\end{tabular}%
}
\end{table*}

%
\textbf{Differential domain:}
\Citet{Arvo1994} derived closed form expressions for irradiance derivatives in polygonal environments, and \citet{Holzschuch1995,Holzschuch1998} derived second-order illumination derivatives for error control in the radiosity algorithm. Local differentials have also proven useful for texture filtering~\citep{Igehy99,Suykens:2001:Path}, photon density estimation~\citep{Schjoth:2007:PD,jarosz11comprehensive}, and spectral rendering~\citep{ElekEGSR2014}. \Citet{ramamoorthi2007} analyzed gradients of various surface lighting effects, including occlusions, and showed how these can be used for adaptive sampling and interpolation in image space.
\Citet{Lehtinen2013sg} and follow-up work~\citep{Manzi2014siga}, proposed to compute image gradients instead of actual luminance values in Metropolis light transport (MLT), and feed a Poisson solver with these gradients to reconstruct the final image. Later work~\citep{Kettunen2015sg,Manzi2015egsr} extended the applicability of this gradient domain idea to simpler Monte Carlo path tracing methods, and demonstrated how solving light transport in the gradient domain improves over primal space, while remaining unbiased. \Citet{rousselle16image} showed how such Poisson-based reconstruction approaches can be directly formulated as control-variate estimators.
%
\Citet{Kaplanyan2013APPM} leveraged second-order derivatives of irradiance to estimate optimal kernel bandwidth in progressive photon mapping, focusing on surface light transport only.

%

%
Closely related to our work,
\citet{Belcour2014} performed a frequency analysis of light fields within participating media. They summarize the local light field using covariance matrices, which provides Hessians of fluence  (up to sign) due to scattering and absorption. Their approach \major{explicitly accounts for radiance changes only in the plane perpendicular to ray propagation, needing to average the per-light-path information from many rays to compute the 3D fluence spectrum.} To account for visibility changes, they also require precomputing the covariance matrices in a finite neighborhood, sacrificing locality and incurring the cost of scene voxelization. In contrast, we provide a fully local method for computing first- and second-order derivatives of media radiance, without requiring voxelization, all while accounting for changes due to visibility, scattering, and transmittance.

\textbf{2D spaces: }
Simplification to lower-dimensional spaces is a recurring tool used in problem analysis. In image synthesis, reduction to hypothetical 2D worlds has been used to obtain insights and illustrate the benefits of more complex 3D approaches~\citep{Heckbert92,Orti1996}. More recent analyses of derivative and frequency domains~\citep{Durand2005,ramamoorthi2007,Mehta2013}, as well as recent work on complex reflectance filtering~\citep{Yan2014Rendering,Yan2016Position} \minor{reduce the complexity of their derivations by performing them in 2D}, before showing how the gained insights generalize to 3D. \Citet{jarosz12theory} introduced a 2D surface radiometry and global illumination framework, and showed how \major{this allows for a more practical analysis of 2D versions of standard rendering algorithms due to faster computation and simpler visualization}. Other fields such as acoustic rendering have recently benefited from 2D reduction to provide interactive simulations~\citep{Allen2015}. 
\minor{   Two-dimensional simulations have also been proved useful to synthesize higher-dimensional light transport, as in transient rendering \citep{bitterli2016virtualfemto,Jarabo2014transient}}. 
%
%
In this paper we follow a similar methodology as \citet{jarosz12theory}, providing a novel 2D radiometry framework for participating media.

\section{2D and 3D Light Transport in Participating Media}
\label{sec:2d_light_transport}
\begin{figure*}[t]
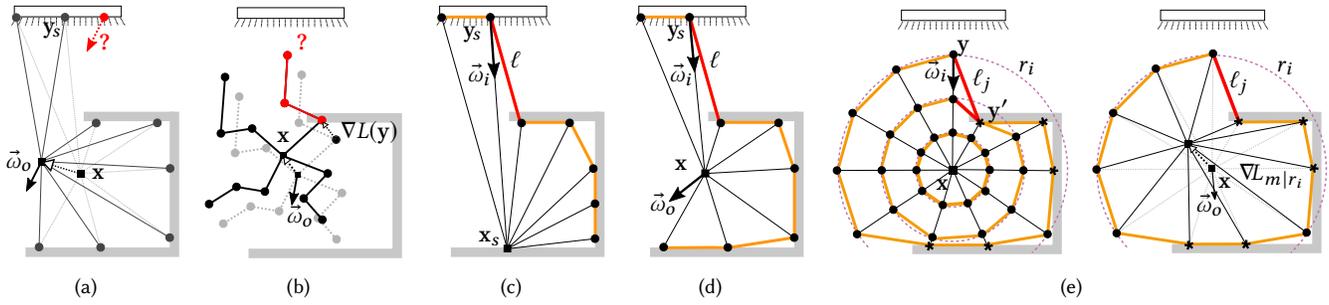

	\subcaptionbox{\label{fig:SSJaroszDerivatives}}{
		\def\svgwidth{0.13\textwidth}
		\input{2DStoM_jarosz.pdf_tex}
	}
	\subcaptionbox{\label{fig:MSDerivativesJarosz}}{
		\def\svgwidth{0.15\textwidth}
		\input{multipleScatteringOccUnawareDerivatives.pdf_tex}
	}
	\subcaptionbox{\label{fig:SurfaceSchwarzhaupt}}{
		\def\svgwidth{0.13\textwidth}
		\input{2DStoS.pdf_tex}
	}
	\subcaptionbox{\label{fig:SSOurDerivatives}}{
		\def\svgwidth{0.13\textwidth}
		\input{2DStoM.pdf_tex}
	}
	\subcaptionbox{\label{fig:MSDerivativesOurs}}{
		\def\svgwidth{0.37\textwidth}
		\input{multipleScatteringOccAwareDerivatives.pdf_tex}
	}
	\caption{%
		\major{\protect\Citeauthor{jarosz08radiance}'s~\protect\citeyear{jarosz08radiance} point-to-point approach for computing first-order derivatives of single (a) and multiple (b) scattering ignores radiance that becomes occluded/disoccluded (red) as $\xm$ is translated. \protect\Citet{schwarzhaupt12practical} compute occlusion-aware derivatives (c) of diffuse surface irradiance by considering the occlusion-free subdivision (orange) of surrounding geometry as seen from $\xs$. We compute occlusion-aware first- and second-order derivatives (d,e) by constructing such occlusion-free subdivisions (orange) of the scene, both at surface locations for single scattering (similar to Schwarzhaupt's work), and also at ray-marched media locations for multiple scattering. Red segments represent approximations of both single and multiple scattering occlusions. Starred points $\star$ in (e) represent black samples at surfaces that occlude radiance from media.}}
	\label{fig:MS}
\end{figure*}
%
%
We describe here the main radiometric aspects of working in a two-dimensional domain, compared to 3D. Similar to \citet{jarosz12theory}, we assume an \textit{intrinsic model} where light is generated, scattered, and absorbed within a plane, thus ensuring energy conservation.




The outgoing radiance at a point $\xm$ in a medium is defined as the angular integral of the incident radiance $\Radi(\xm,\omegain)$, modulated by the scattering phase function $\pfs(\xm, \omegain, \omegaout)$:
%
\begin{align}
%
\Rado (\xm,\omegaout) &= \int_{\Omega} \pfs(\xm, \omegain, \omegaout) \, \Radi(\xm,\omegain) \;\diff\omegain ,
\label{eq:outLF}
\end{align}
where $\omegain$ and $\omegaout$ are \minor{directions over the spherical domain $\Omega$ pointing into and out of the point $\xm$ respectively}.
%
The incident radiance $\Radi = L_m + L_s$ is the sum of radiance arriving from the surrounding medium ($L_m$) and from surfaces ($L_s$):
\begin{align}
\label{eq:mediaLFin}
\Rad_m (\xm,\omegain) &=\! \int_0^{s}\! \sca(\ym(t)) \, \Tr (\xm, \ym(t)) \, \Rado (\ym(t),\omegain)\;\diff t ,
\\
\label{eq:curveLFin}
\Rad_s (\xm,\omegain) &= \Tr (\xm, \ys) \, \Rad_o (\ys,\omegain),
\end{align}
where
$\ym(t) = \xm - t \omegain$ is a point in the medium, and $\ys$ is a point on a surface at distance $s$ 
with outgoing radiance \minor{$\Rad_o$} modeled by the rendering equation~\citep{Kajiya1986}.
The transmittance $\Tr$ models the attenuation due to scattering and absorption between two points, and \major{$\sca(\xm)=\rho\crossSection_s$ is the scattering coefficient at $\xm$, with $\rho$ and $\sigma_s$ the density and scattering cross-section in the medium, respectively}. We detail our notation in \autoref{tab:notation}, and highlight the main radiometric differences between self-contained 2D and 3D worlds, described below.

%
\textbf{Differences in 2D: } When moving to a 2D world, the intrinsic radiometric model implies that all radiance travels within a planar medium, scattering therefore over angle instead of solid angle.
This means that radiance falls off with the inverse distance instead of inverse squared distance~\citep{jarosz12theory}; this will become important in our analysis of first- and second-order derivatives.

The main changes when applying \EqsRange{outLF}{curveLFin} in 2D are:
\begin{itemize}
	\item The integration domain $\Omega$ of \Eq{curveLFin} becomes circular instead of spherical.
	\item The phase functions in 2D must be normalized over the circle, not the sphere, of incident directions.
	\item \minor{$\Rad_o (\ys)$} now indicates radiance from the closest curve (the 2D equivalent of a 3D surface).
\end{itemize}
%
%
In the next sections, we use this self-contained 2D world to better depict and reason about the improvements of our new occlusion-aware gradients and Hessians for media (\Sec{inscatDerivatives}), and our second-order error metric (\Sec{errorcontrol}), before extending them to a more practical three-dimensional world.
Working in 2D also allows us to avoid collapsing a 3D scene into a 2D image for visualization, where information from many media points would contribute to a single image pixel. This allows us to illustrate the performance of our algorithm in a more intuitive way (\Sec{results}) and to depict the introduced errors more clearly.

\subsection{Radiance Caching in Participating Media}
\label{sec:rcbackground}
\begin{figure}[t]
	\centering
	\includegraphics[width=\columnwidth]{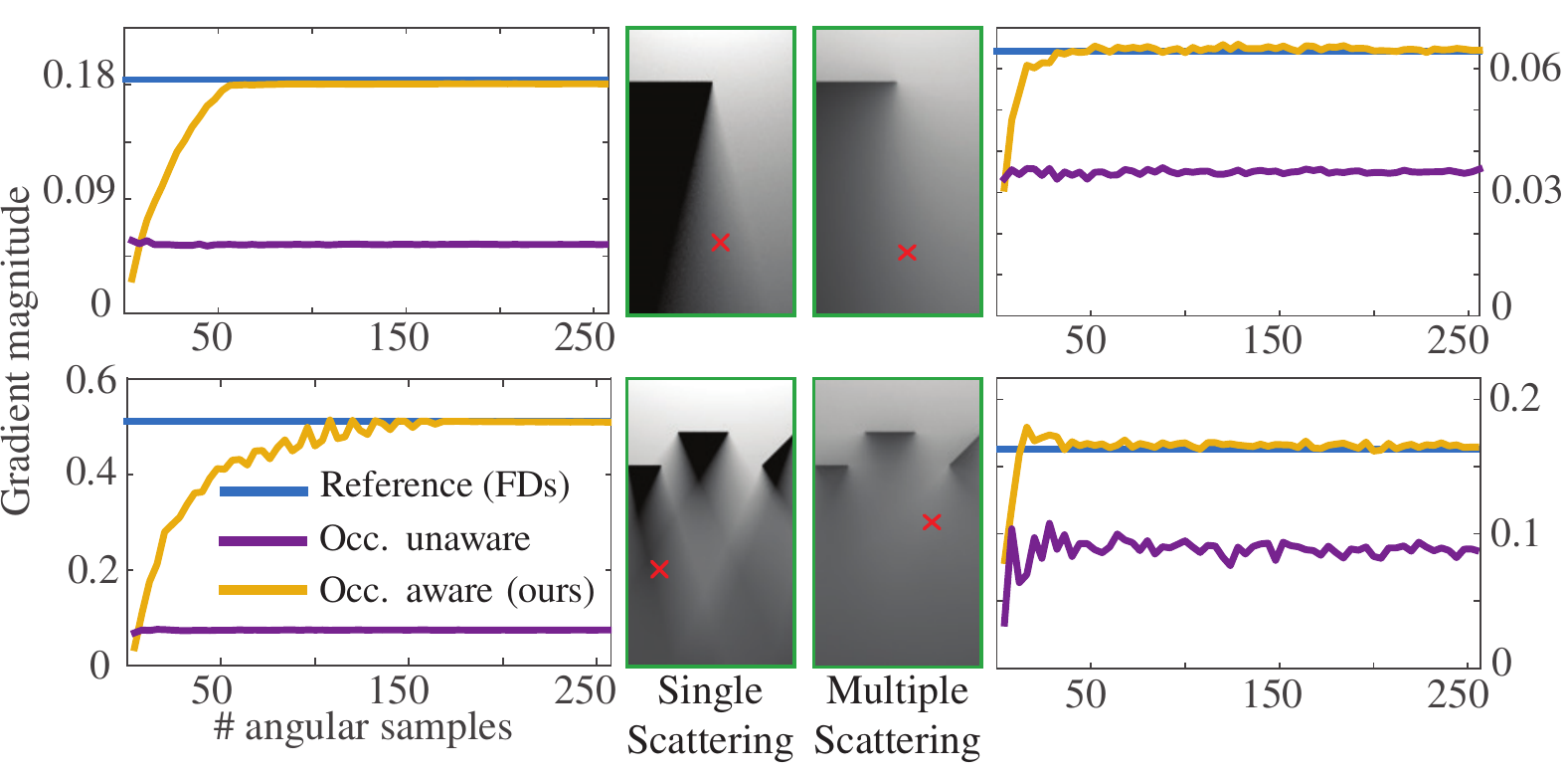}
	\caption{\major{Compared to prior occlusion-unaware gradients (purple), our gradients (yellow) converge to the reference solution (blue) with increasing angular sample count both for single scattering (left) and multiple scattering (right). \major{The convergence plots are computed in the red crosses in the respective middle images.} }
	%
	%
	%
	}
	\label{fig:2DSSconvergence}
\end{figure}

\major{Before deriving our second-order, occlusion-aware volumetric radiance caching approach, we first summarize \citeauthor{jarosz08radiance}'s~\citeyear{jarosz08radiance} original formulation. To determine the radiance at any point $\xpr$ in the medium\footnote{Throughout the text, $\xpr$ represents points where we approximate radiance by interpolating the cache points, while $\xm$ represents points where we compute radiance and its derivatives explicitly.}, their algorithm first tries to approximate this value by extrapolating (in the log domain) the cached radiance $\Rad_k$ from nearby cache point locations $\xm_k$ along their respective gradients:
\begin{align}
\Rado(\xpr,\omegaout) &\approx \exp \left[\frac{\sum_{k\in C} \left(\ln{\RadoIdx{k}} + \GRx\ln{\RadoIdx{k}}\cdot \Deltaxpr \right)\mathrm{w}(\xm_k, \xpr)}{\sum_{k\in C} \mathrm{w}(\xm_k, \xpr)}\right],
\label{eq:jaroszExtrapolation}
\end{align}
with $\Deltaxpr=(\xpr - \xm_k)$.
Here $\GRx \ln \RadoIdx{k} = \GRx\RadoIdx{k} / \RadoIdx{k}$ is the log-space translational gradient of cache point $\xm_k$, and $\mathrm{w}(\xm_k, \xpr)$ is a weighting function that diminishes the influence of a cache point to zero as $\xpr$ approaches the cache point's valid radius. The collection of nearby cache points $C$ consists of all cache points whose valid radii contain $\xpr$. If no nearby cache points are found, then the algorithm computes radiance using Monte Carlo sampling and inserts the value and its gradient into the cache for future reuse.

\Citet{jarosz08radiance} proposed to compute the valid radii using a metric based on the local log-space radiance gradient:
\begin{align}
R &= \varepsilon \frac{\sum \RadoIdx{j}}{\sum \norm{\GRx \RadoIdx{j}}},
\label{eq:jaroszRadius}
\end{align}
where $\varepsilon$ is a global error tolerance parameter and $\RadoIdx{j}$ and $\GRx \RadoIdx{j}$ are the individual Monte Carlo samples of radiance and translational gradient respectively. \minor{Unfortunately, this error metric is an ad-hoc approximation of the error in the log-scale interpolation, which can lead to difficulty predicting the error in the sample distribution and suboptimal cache distributions. }

\Citeauthor{jarosz08radiance} maintain a separate cache for single/surface scattering and multiple scattering. They compute single-scattering gradients by Monte Carlo sampling the first translational derivative of \Eqs{outLF}{curveLFin} in surface-area form. They trace out many rays in the sphere of directions around point $\xm$ to obtain a number of surface hit points $\ys$. Their gradient calculation, in essence, considers how the radiance $L_s$ from each of these hit points would change (due to changes of transmittance and geometry terms, but not visibility) as $\xm$ translates, but the surface hit points $\ys$ remain fixed (see \Fig{SSJaroszDerivatives}).
For multiple-scattering gradients, they Monte Carlo sample the first derivative of \Eqs{outLF}{mediaLFin}, where the whole set of sampled paths is assumed to move rigidly (see \Fig{MSDerivativesJarosz}), accounting for translational derivatives at each scattering vertex.

This gradient formulation can efficiently compute the local change in radiance of any single Monte Carlo sample, but---by operating independently on each radiance sample---it is not able to capture \emph{global} effects such as visibility gradients. As a consequence, changes in radiance that becomes occluded/unoccluded as the shaded point is translated are not taken into account (see \Figs{SSJaroszDerivatives}{MSDerivativesJarosz}, red).}
As an illustrative example, \Fig{2DSSconvergence} shows how ignoring occlusions (purple line) leads to incorrect single- and multiple-scattering gradients in the penumbra region beneath the occluder.

In the remainder of this paper we describe our novel Hessian-based radiance caching method for participating media that overcomes the aforementioned limitations. In \Sec{inscatDerivatives} we introduce our approach for computing occlusion-aware first- and second-order derivatives of media radiance. Then, in \Sec{errorcontrol} we introduce our Hessian-based error metric and extrapolation method for volumetric radiance caching.

\section{Radiometric Derivatives in Media}
\label{sec:inscatDerivatives}

Following the work of \citet{schwarzhaupt12practical} on global illumination on surfaces, we formulate the radiance at $\xm$ as a piecewise linear representation of the incoming radiance. 
Conceptually, \major{we build an approximated coarse representation of the scene as seen from the media point $\xm$ by triangulating adjacent stochastic angular samples $\ys$ (see \Figs{SurfaceSchwarzhaupt}{SSOurDerivatives})}.
The interesting property of this triangulation is that the geometry term for each triangle (segments in 2D) models the attenuation due to the solid angle; as a consequence, changes in the geometry term (due to translation of $\xm$) model changes in the observed radiance.

We extend \citeauthor{schwarzhaupt12practical}'s~\citeyear{schwarzhaupt12practical} formulation to handle not only light transport from surfaces, but also from media. In the case of surfaces, the sample points $\ys$ are located at the first surface point as seen from $\xm$ in direction $\yx$ (\Fig{SSOurDerivatives}). For points in a participating medium, however, radiance arrives from multiple distances along each direction. We therefore consider a set of concentric triangulations at increasing distances $\sDistance_i$, each representing the outgoing radiance at that particular distance in the medium. If occluding geometry exists closer than the distance $\sDistance_i$, we place a zero-radiance sample at the surface intersection (points marked with $\mathbf{\star}$ in \Fig{MSDerivativesOurs}).
%
\begin{figure}[t]
\centering
\def\svgwidth{.8\columnwidth}
\input{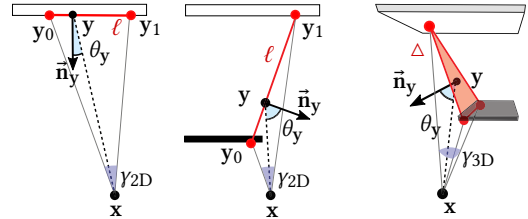}
\caption{Left and center: Visible and occluded cases for 2D surface-media radiance for an angle $\gamma$. Red segment represent the piecewise-linear construction as seen from $\xm$. {Right: 3D interpretation, where occlusions are represented by slanted triangles, and visibility changes are modeled as changes in the 3D geometry term between the triangle points $\yw \in \tri$, and $\xm$.}
}
\label{fig:singlepiece}
\end{figure}

\textbf{Handling Occlusions and Transmittance:}
In essence, \major{we are approximating the integration along $\Omega$, by transforming the scene into a discrete set of \emph{virtual} piecewise linear representations of the geometry and media around $\xm$}. As noted by \citeauthor{schwarzhaupt12practical}, this representation implicitly encodes changes in visibility by means of the geometry term. Our approach for media, however, requires taking transmittance into account and using different geometry terms (see \Fig{SurfaceSchwarzhaupt}), since surface-medium light transport only has a cosine term at the source $\ys$.
%
%
\major{We illustrate this with a 2D example in} \Fig{singlepiece}, left and center: Assuming a constant angle $\gamma$ between vectors $\vv{\xm\y_0}$ and $\vv{\xm\y_1}$, occlusions generate segments $\seg= \y_1 - \y_0$ at grazing angles, with derivatives proportional to the steepness of the segment.
%
When moving within the medium, the projected angle of $\seg$ towards $\xm$ is proportional to $\cosy$, and therefore the radiance from $\seg$ increases with $\cosy$. This allows modeling the visibility changes as a  change on the 2D geometry term $\Geom = \cosy/\norm{\xy}$.
This principle holds also for 3D, as \Fig{singlepiece}, right, shows: Occlusions are represented by slanted triangular faces, and visibility changes are modeled as changes in the 3D geometry term between the triangle points $\yw \in \tri$ and $\xm$.
We leverage this equivalence to provide a unified formulation for radiance derivatives, applicable both to 2D and 3D\footnote{For convenience, we formulate all the equations in terms of 2D media and geometry subdivisions in segments~$\seg$, but all formulae are equally applicable in 3D by substituting segments $\seg$ by triangles $\tri$.}.

%
%

Using the formulation presented before, we approximate $\Rado(\xm,\omegaout)$ by discretizing the space into a set of concentric rings $\Rings$ as:
\begin{align}
\Rado(\xm,\omegaout)
\approx \sum_{\ring_i\in \Rings}\frac{1}{\pdf(\ring_i)}\sum_{\segj\in \Segments_i}\RadoIdx{j}(\xm,\omegaout),
\label{eq:ApprxRadiance}
\end{align}
%
%
where the last ring $\ring_s \in \Rings$ has all its vertices on surfaces, $\Segments_i$ is the set of segments for ring $\ring_i$, and $\pdf(\ring_i)$ is the probability of sampling a particular distance when building the ring (for the surface ring, we have $\pdf(\ring_s)=1$). $\RadoIdx{j}$ is the radiance contributed by each segment $\segj \in \Segments_i$, defined by the integral:
\begin{align}
\!\RadoIdx{j}(\xm,\omegaout) = \int_{\segj} \!\! \pf(\xm,\omegain,\omegaout) \, \GeomMs(\xm,\yw) \, \Tr(\xm,\yw) \, \Rado(\yw,\omegain)\; \diff \yw.\!
\label{eq:SegmentRadiance}
\end{align}
By construction, the visibility between $\xm$ and $\yw$ is ${\Vis(\xm,\yw)=1}$, and $\yw$ is a point on a \emph{virtual} surface; 
we thus need to account for the foreshortening at $\yw$. This allows for a unified formulation of both surface-to-medium and medium-to-medium radiance derivatives, using the same geometry term in both cases. \major{Note that we have merged together the phase function $\pfs(\xm,\omegain,\omegaout)$ and scattering coefficient $\sca(\xm)$ as a directional scattering function $\pf(\xm,\omegain,\omegaout)=\sca(\xm)\pfs(\xm,\omegain,\omegaout)$, to make the following derivations simpler.}

Differentiating \Eq{ApprxRadiance} with respect to $\xm$ provides approximations for the first and second order derivatives:
\begin{align}
\label{eq:SegmentRadianceGradient}
\GRx \Rado(\xm,\omegaout)
&\approx \sum_{\ring_i\in \Rings}\sum_{\segj\in \Segments_i}  \frac{\GRx \RadoIdx{j} (\xm,\omegaout)}{\pdf(\ring_i)},\\
\HSx \Rado (\xm,\omegaout)
&\approx \sum_{\ring_i\in \Rings}\sum_{\segj\in \Segments_i}  \frac{\HSx \RadoIdx{j} (\xm,\omegaout)}{\pdf(\ring_i)},
\label{eq:SegmentRadianceHessian}
\end{align}
which in turn require differentiating the radiance from each segment.
%

\begin{figure*}[th]
	\centering
	\includegraphics[width=0.97\textwidth]{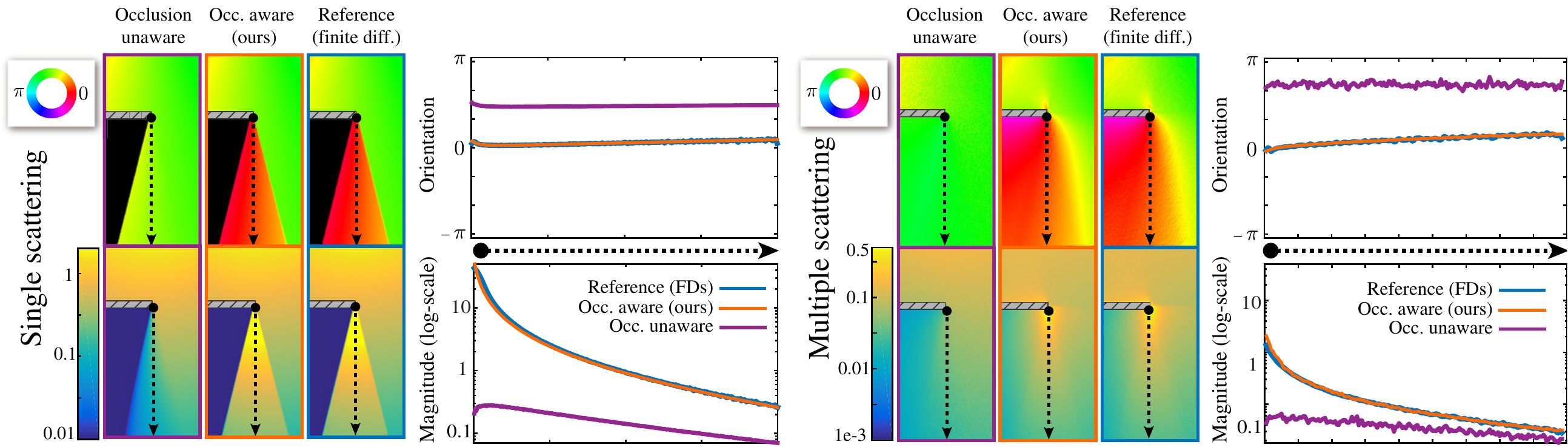}
	\caption{2D single and multiple scattering gradients in similar setup to \Fig{2DSSconvergence}, top.
		Compared against an occlusion-unaware reference solution~\protect\cite{jarosz08radiance}, our method correctly captures both gradients orientation (color-coded angle), and magnitude. The graphs show the evolution of the gradient across the dotted black line for both methods (purple, orange), and the reference solution (blue).
	}
	\label{fig:2Dgradients}
\end{figure*}

Unfortunately, we cannot compute \Eq{SegmentRadiance} and its derivatives analytically in closed-form, while computing it numerically would be prohibitively expensive. We instead introduce a set of assumptions to build a closed-form approximation:

%
%
%
\begin{itemize}
\item For a sufficiently fine subdivision the angle $\gamma$ tends to 0, so $\win$ can be regarded as constant for the whole segment, and \major{$\pf(\xm, \win, \wout) = \pf(\xm,\omegal,\wout)$, with $\omegal$ a fixed direction from $\xm$ to a point in segment $\seg$}. 
\item For all $\yw \in \seg$, we assume constant $\Tr(\xm, \yw) = \Tr(\xm, \xl) $, and $\Rado(\yw,\omegain) = \Rado(\xl, \win)$. 
Following existing approaches for surface irradiance, we choose $\xl$ as the furthest point in the segment $\seg$, which will be the first to be occluded/unoccluded.
\end{itemize}
These assumptions allow us to significantly simplify the integral in \Eq{SegmentRadiance} to:
\begin{align}
\RadoIdx{j}(\xm,\omegaout) &\approx \pf(\xm,\omegalj,\omegaout) \, \Tr(\xm,\xlj) \, \Rado(\xlj,\omegain)\!\int_{\segj}\!\GeomMs(\xm,\yw) \; \diff \yw \nonumber \\
& = \pf(\xm,\omegalj,\omegaout) \, \Tr(\xm,\xlj) \, \Rado(\xlj,\omegain) \, \FFDDMj(\xm),
\label{eq:ApproxSegmentRadiance}
\end{align}
which now admits a closed-form solution in both 2D and 3D (see \Apxx{segMediaFormFactor2D}{fluence3Dsegment}). More importantly, this allows us to approximate the derivatives of $\RadoIdx{j}$ in closed form as:
\major{
\begin{align}
\GRx \RadoIdx{j} &\approx \Rad_F \GRx \pf + \GRx \Rad_F \pf,
\label{eq:SSSegALGradient}\\
\HSx \RadoIdx{j} &\approx \Rad_F \HSx \pf + \GRx \Rad_F \GRxT \pf + \GRx\pf \GRxT\Rad_F + \HSx \Rad_F  \pf,
\label{eq:SSSegALHessian}\\
\intertext{where }
\GRx \Rad_F &= \redRad \GRx \FFDDM +  \GRx \redRad \FFDDM,\\
\HSx \Rad_F &= \redRad \HSx \FFDDM +  \GRx \redRad \GRxT \FFDDM +  \GRx \FFDDM \GRxT \redRad + \HSx \redRad \FFDDM, \\
\label{eq:scaledRadianceDerivatives}
\GRx \redRad &= \Rado \GRx  \Tr + \GRx \Rado \Tr , \\
\HSx \redRad & = \Rado \HSx \Tr +  \GRx \Rado \GRxT \Tr +  \GRx  \Tr \GRxT \Rado + \HSx \Rado \Tr.
\label{eq:reducedRadianceDerivatives}
\end{align}
}%
For brevity we have omitted function parameters, and we express gradients and Hessians in terms of the scaled radiance $\Rad_F = \FFDDM \redRad$, and the reduced radiance $\redRad = \Rado \Tr$. \major{While \EqsRange{ApproxSegmentRadiance}{reducedRadianceDerivatives} are general, we restrict our work to  Lambertian surfaces and isotropic, homogeneous media (in \Sec{discussion} we discuss how to extend it to anisotropic and heterogeneous media). This means that both $\Rado$ and $\pf$ are constant, and therefore their derivatives cancel out as $\GRx\Rado=\HSx\Rado=\GRx\pf=\HSx\pf=0$, removing directional dependences; this allows us to simplify \Eqs{SSSegALGradient}{SSSegALHessian} to:
\begin{align}
&\GRx \RadoIdx{j} \approx \Rado\pf\left( \Tr \GRx \FFDDM +  \GRx  \Tr \FFDDM\right) ,
\label{eq:SSSegALGradientSimp}\\
&\HSx \RadoIdx{j} \approx \Rado\pf\left(\Tr \HSx \FFDDM + \GRx  \Tr \GRxT \FFDDM +  \GRx \FFDDM \GRxT  \Tr + \HSx  \Tr \FFDDM \right).
\label{eq:SSSegALHessianSimp}
\end{align}
} We refer to Appendices \ref{ap:transmittance}, \ref{ap:segMediaFormFactor2D} and \ref{ap:fluence3Dsegment} for all the terms.

\major{By construction, our formulation in \Eq{ApprxRadiance} and its derivatives (\Eqs{SegmentRadianceGradient}{SegmentRadianceHessian}) are biased but consistent estimators of $\Rado(\xm,\omegaout)$, $\GRx\Rado(\xm,\omegaout)$, and $\HSx\Rado(\xm,\omegaout)$, respectively. In addition the assumptions imposed in \Eq{ApproxSegmentRadiance} introduce some additional bias due to the piecewise assumption in the scattering $\pf$, transmittance $\Tr$, and radiance terms $\Rado$. However, as shown in \Fig{2DSSconvergence} our formulation converges accurately to the actual derivatives.}
Note that we use this biased but consistent approximation only to compute first- and second-order derivatives of media radiance (\Eqs{SegmentRadianceGradient}{SegmentRadianceHessian}), while computing actual radiance values (\Eq{outLF}) using the standard unbiased Monte Carlo estimator.
\major{In the following, we describe how to use the derivatives in \Eqs{SegmentRadianceGradient}{SegmentRadianceHessian} for interpolating radiance from a set of cache points, and define an error metric for such interpolation.}

\section{Second-order error control for media radiance extrapolation}
\label{sec:errorcontrol}
The error in radiance caching is controlled by a tolerance value $\sErrorTolerance$, and depends both on how radiance is extrapolated, and on the radiance moments at cache point $\xm$. These moments define a valid bounding region $\VRegion$ where a point $\xpr$ can be used for extrapolation.
We provide \minor{here the key ideas and resulting equations for the valid regions in the context of 2D and 3D participating media and provide detailed derivations in the supplementary material}.

Existing work on radiance caching for participating media estimates the relative error using radiance gradients at \xm.
However, ignoring higher-order derivatives creates suboptimal cache distributions that often oversample regions near surfaces and light sources.
Given the radiance and the first $n$ derivatives at a media point \xm, we can approximate radiance at point $\xpr \in \VRegion$ using an $n\th$-order Taylor expansion.
\major{Following previous work~\citep{schwarzhaupt12practical} we truncate to order one, approximating $\Rado(\xpr,\omegaout)$ as:
\begin{align}
\Rado(\xpr,\omegaout) \approx \Rado(\xm,\omegaout) + \GRx \Rado(\xm,\omegaout) \Deltaxpr.
\label{eq:extrapolationTaylor}
\end{align}
\major{Since we focus on isotropic media, we remove the directional dependence in the following derivations to simplify notation.}
By using a second order expansion of $\Rado(\xm)$ as our oracle, we can approximate the relative error $\estRelError(\xpr)$ of the extrapolation as:
\begin{align}
\estRelError(\xpr)
& \approx \frac{\left|\DeltaxprT \HSx \Rado(\xm) \Deltaxpr\right|}{2\,\Rado(\xm)},
\label{eq:errorTaylor}
\end{align}
with $\HSx \Rado(\xm)$ the Hessian matrix of $\Rado(\xm)$.
}
This expression is similar to the second-order error metric proposed by~\citet{jarosz12theory} and follow-up work by~\citet{schwarzhaupt12practical}, although these works dealt with surfaces only.

\major{By integrating \Eq{errorTaylor} in the neighborhood of $\xm$ for a given error threshold $\sErrorTolerance$, we can express the valid region in two-dimensional media as an ellipse with principal radii $R^{\lambda_i}_{\DD}$ (see Equations (S.9)-(S.12) in the supplemental for the complete derivation): }
\begin{align}
R_{\DD}^{\lambda_i}  =\sqrt[4]{\frac{4 \Rado(\xm) \sErrorTolerance}{\pi|\lambda_i|}},
\label{eq:radiusAniso2D}
\end{align}
%
%
where $\lambda_i$ is the $i$-th eigenvalue of the radiance Hessian $\HSx \Rado(\xm)$. This formula is analogous to the relative error metric presented by Schwarzhaupt and colleagues~\citeyear{schwarzhaupt12practical} for surfaces, but here the radii are computed by taking the principal components of the \emph{volumetric} radiance Hessian. Adding the third dimension, the valid region for a cache point becomes a 3D ellipsoid, whose principal radii are:
%
\begin{align}
R_{\DDD}^{\lambda_i}  =\sqrt[5]{\frac{15 \Rado(\xm) \sErrorTolerance}{4\pi|\lambda_i|}}.
\label{eq:radiusAniso3D}
\end{align}
%

Our second-order error metric and its derived radius assume knowledge of the radiance and its derivatives at $\xm$. In practice, these are usually computed by Monte Carlo techniques, which lead to other sources of error such as variance (inherent to Monte Carlo sampling), or bias (due to inaccuracies computing the derivatives). 

\major{The presented metric describes the error introduced by extrapolation from a single cache point in participating media. However, at render time, we compute radiance at each shaded point by interpolating from \emph{multiple} cached points, as:
\begin{align}
\Rado(\xpr) &\approx \frac{\sum_{k\in C}^{} \left[\Rado(\xm_k) + \GRx \Rado(\xm_k) \cdot\Deltaxpr\right]\mathrm{w}(\xm_k, \xpr)}{\sum^{}_{k\in C}\mathrm{w}(\xm_k, \xpr)},
\label{eq:radianceTaylor}
\end{align}
with $C$ the set of cache points whose radii include $\xpr$, and $\mathrm{w}(\xm_k, \xpr)$ the interpolation kernel. Following \citet{jarosz08radiance}, we use a cubic interpolation kernel  $\mathrm{w}(\xm_k, \xpr) = 3d^2-2d^3$ with ${d\!=\!1\!-\!\|\xpr\!-\xm_k\|\,R_k^{-1}}$.}
\minor{Since Equation~\eqref{eq:radianceTaylor} only interpolates from cache points which predict a maximum error $\estRelError<\sErrorTolerance$ at \xpr, the error of the weighted sum is equally upper-bounded by $\sErrorTolerance$.}
\major{Note that, as opposed to \citet{jarosz08radiance}~\eqref{eq:jaroszExtrapolation}, we interpolate in linear space, where the error is more accurately predicted by our Hessian-based metric described in Equation~\eqref{eq:errorTaylor}.
}
\section{Results}
\label{sec:results}

In the following we illustrate the accuracy and benefits of our method. %
%
We start showing our results in a two-dimensional world, and compare it against a 2D version of the current state-of-the-art method~\cite{jarosz08radiance}. We refer the reader to the supplementary material for the additional expressions to compute two-dimensional occlusion-unaware gradients.
Then, we move to 3D, to demonstrate that our results are also consistent in a more practical three-dimensional scenario. For comparison purposes, all 3D insets show only single and multiple scattering in media, discarding surface radiance. \major{Unless it is explicitly mentioned, we use isotropic points with the smallest principal axis of the Hessian. This is the most costly scenario for our method in comparison to previous work, since we cannot adapt to the signal as faithfully as with anisotropic points, and therefore require more points. }

\textbf{Implementation:}
\major{We compute both radiance and derivatives at point $\xm$ by stratified sampling uniformly in the sphere, with equal solid angle strata (in the case of 2D, this stratification is in the circle, using equal angle stratification). This reduces variance compared to pure uniform sampling. More importantly, it allows to very simply \minor{build the subdivision using the angular samples}, by just connecting samples from adjacent strata~\citep{schwarzhaupt12practical}. This stratification is used for both media and surfaces, including area light sources, while other direct light sources (such as directional or point lights) must be handled separately. 
The accuracy of the \minor{subdivision} for computing the derivatives relies on a dense sampling of the angular domain, and, as in any sampling problem, \minor{our sampling rate limits the amount of radiance changes that we can recover.}
This is especially important when capturing fine details such as small light sources, which are not computed using next event estimation (NEE), but could also be important in high-frequency fluctuations of radiance in media. However, in practice our method presents much better convergence than previous work~\citep{jarosz08radiance} with increasing number of angular samples, as shown in \Fig{2DSSconvergence}. Introducing a NEE-aware \minor{subdivision} combined with the standard angular one via multiple importance sampling could significantly improve the performance of the \minor{derivative computation, although we leave this to future work}. 
\minor{We perform the subdivision within the medium by uniformly ray-marching the medium at discrete distances around $\xm$, and joining adjacent angular samples within each marching step (see \Fig{MSDerivativesOurs})}.
}

Unless stated otherwise, single scattering in all compared methods refers to radiance emitted or reflected (first bounce) by surfaces. We limit multiple scattering to the second bounce for all methods. We did this mainly to reduce excessive variance when computing reference derivatives with finite differences. Note that both occlusion-unaware and occlusion-aware methods are equally applicable to higher number of media bounces, although they usually require a high number of samples to obtain noiseless solutions. 

\major{Following previous methods~\citep{jarosz08radiance}, we first pre-populate the cache by uniformly sampling a ray from the camera, and ray-marching along the media, placing cache points in case they do not fulfill our error metric (\Sec{errorcontrol}). \minor{At render time, we evaluate \Eq{radianceTaylor} at ray-marched points $\xm'$ in the medium, extrapolating radiance from the surrounding valid cache points. If no valid cache points are found for $\xm'$ then we compute its radiance and derivatives, and add it to the cache.} As in previous methods~\citep{jarosz08radiance}, we separate single and multiple scattering caches, each in a different octree for efficient cache query. }

%
All results were computed on a desktop PC with an Intel Core i7 3.4 GHz CPU and 16GB RAM. \major{Note that all methods used for rendering comparisons of the complex 3D scenes \emph{Whiteroom} and \emph{Staircase} were accelerated with Embree ray-tracing kernels \cite{wald2014embree}, and therefore the performance with respect to the other 3D scenes is higher. }


%
\subsection{Results in 2D}
\begin{figure}[t]
	\centering
	\includegraphics[width=\columnwidth]{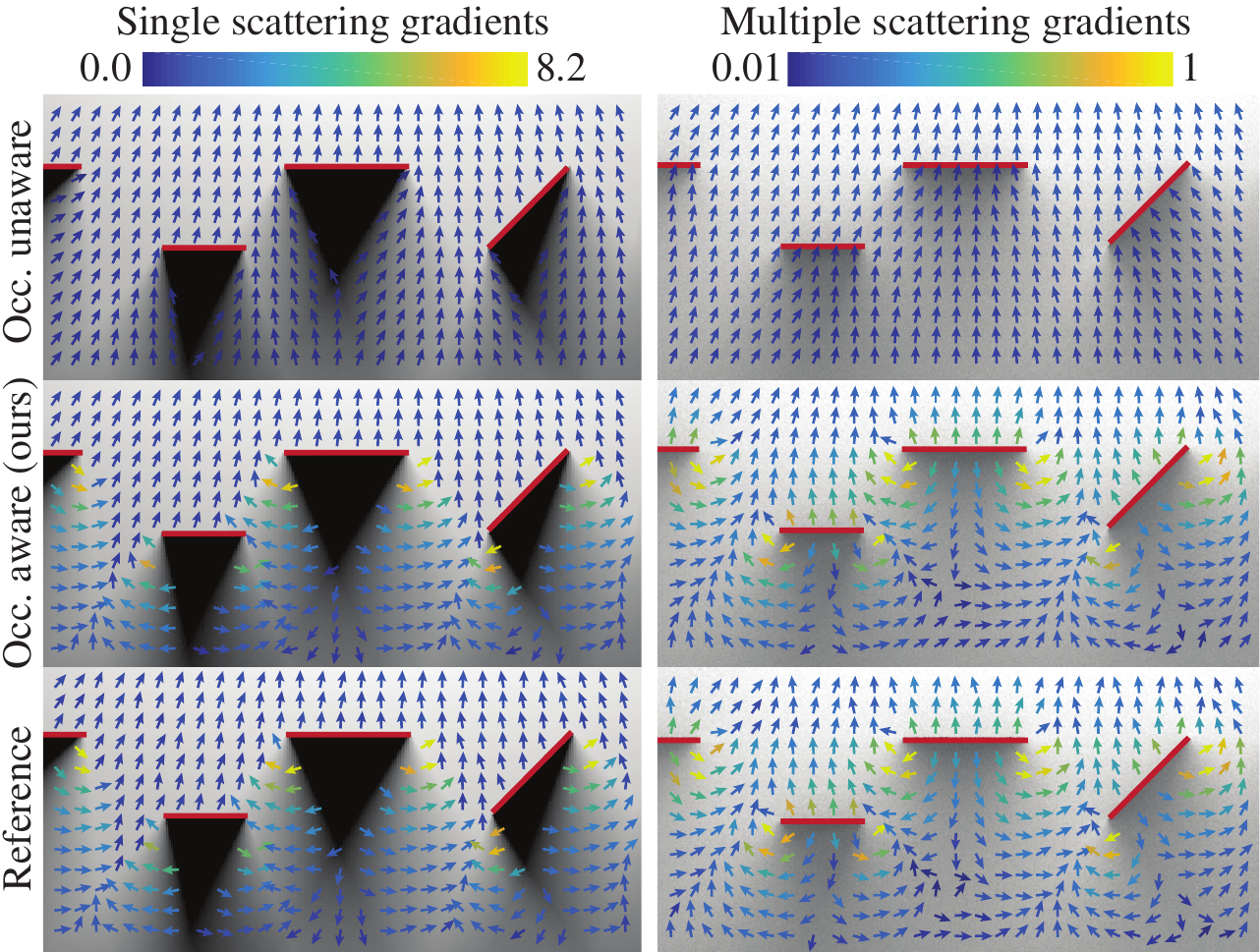}
	\caption{Radiance gradients at discrete locations in 2D computed with occlusion unaware, and our occlusion aware methods, compared against reference gradients (bottom row) computed with path traced finite differences using 4M samples/gradient. Left column, top and middle, show single scattering gradients computed with 256 angular samples/gradient. Right column, top and middle, show multiple scattering gradients computed with 65536 samples/gradient (256 angular $\times$ 256 ray samples).}
	\label{fig:2DGradientsGrid}
\end{figure}
%


To evaluate the error introduced by our occlusion-aware computations of derivatives in a clear, intuitive way, we rely on their two-dimensional versions.
%
%
In \Fig{2DSSconvergence} we showed the convergence of gradient computation with the number of angular samples. Previous approaches not taking into account visibility changes fail to estimate the gradient.
In contrast, our derivative formulation converges to the actual gradient, even in areas of penumbra for both single and multiple scattering. The quality of our estimated derivatives increases with the number of angular samples, since the approximations introduced by our assumptions vanish as the strata size diminishes.

In \Fig{2Dgradients} we compare the evolution of single and multiple scattering gradients across a penumbra region, computed with our method and previous work. We illustrate them in polar coordinates (magnitude and orientation) in a simple scene with a medium illuminated by an area light on top, and a line acting as an occluder within the medium. We compute reference gradients with path traced finite differences. Our approach manages to correctly compute both gradients magnitude and orientation in the penumbra region. The right graphs show a progression of gradients along the dotted line. The graphs show that our method is able to match the ground-truth, while the occlusion-unaware method both underestimates the magnitude of the gradient and computes an incorrect direction.

\Fig{2DGradientsGrid} shows a comparison of gradients (shown as a vector field) with the occlusion-unaware method, our technique, and a ground-truth solution computed with finite differences.
%
\begin{figure}[t]
	\centering
	\includegraphics[width=\columnwidth]{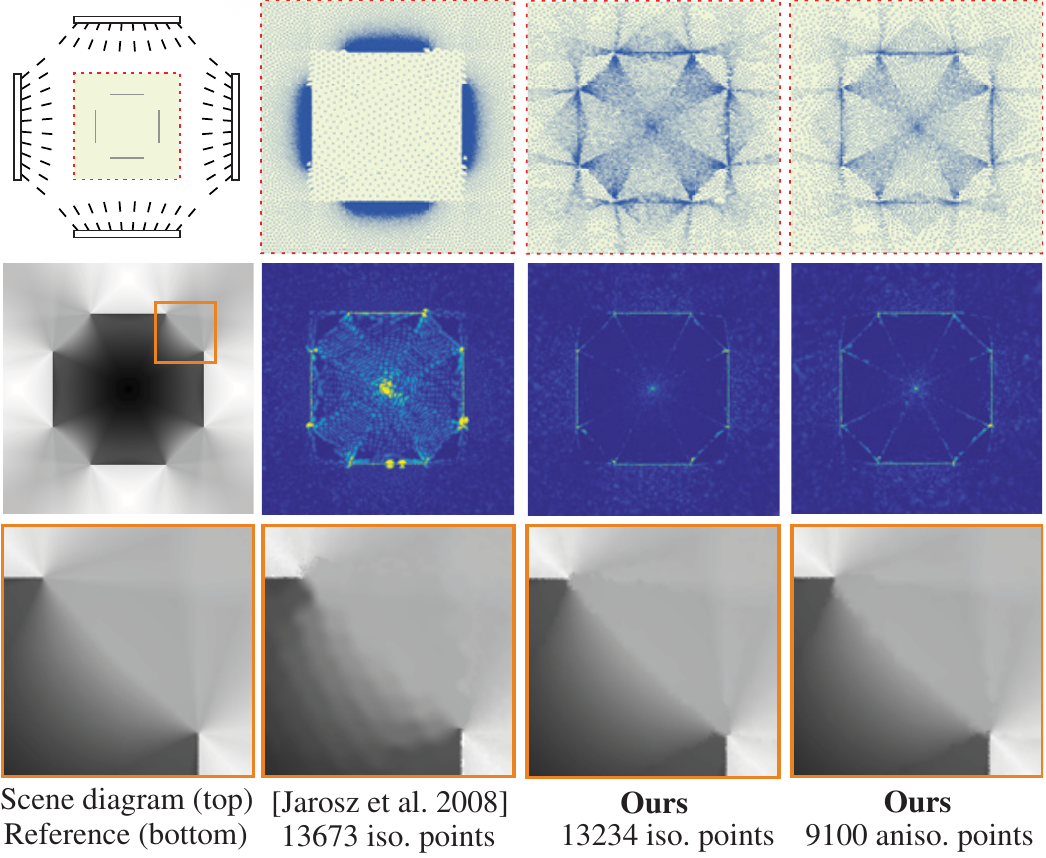}
	\caption{Single scattering in a 2D setup with four line lights and four occluders. Point distributions (top row) show how a occlusion-unaware gradient metric~\protect\cite{jarosz08radiance} fails to estimate the correct radiance changes in complex shadows, while tending to concentrate cache points near reflecting geometry. In contrast, our algorithm distributes points according to occlusion-aware, second-order derivatives of radiance, capturing complex light patterns more accurately. Leveraging curvature information in the Hessians enables anisotropic cache points that further reduce the number of required cache points while maintaining quality (see error maps).}
	\label{fig:crossShadows}
\end{figure}
Our method correctly captures complex radiance changes, including strong changes near occluder boundaries, closely matching the ground-truth reference.

Our error metric takes into account second-order derivatives to drive sample-point density in the scene. Since we use the estimated occlusion-aware Hessians as an oracle of the error, this allows us to place more cache points in areas with higher frequencies. Additionally, our improved gradients allow for a more accurate extrapolation within the valid region of the cache points.
\Fig{crossShadows} (top) shows a scene with overlapping shadows, created by four lights and four occluders (top-left diagram indicates the shaded region in green). Previous work (second column) drives point density based on the log-space gradient of radiance; in practice this tends to drastically increase point density near light-reflecting geometry, failing to efficiently sample shadowed regions.
This can only be mitigated by radius-clamping heuristics (in this case based on the pixel size), thus breaking the principled properties of the approach. In contrast, our method (last two columns) does not rely on heuristics and manages to correctly capture shadows by placing more points near shadow boundaries.

\begin{figure}[t]
	\centering
	\includegraphics[width=0.85\columnwidth]{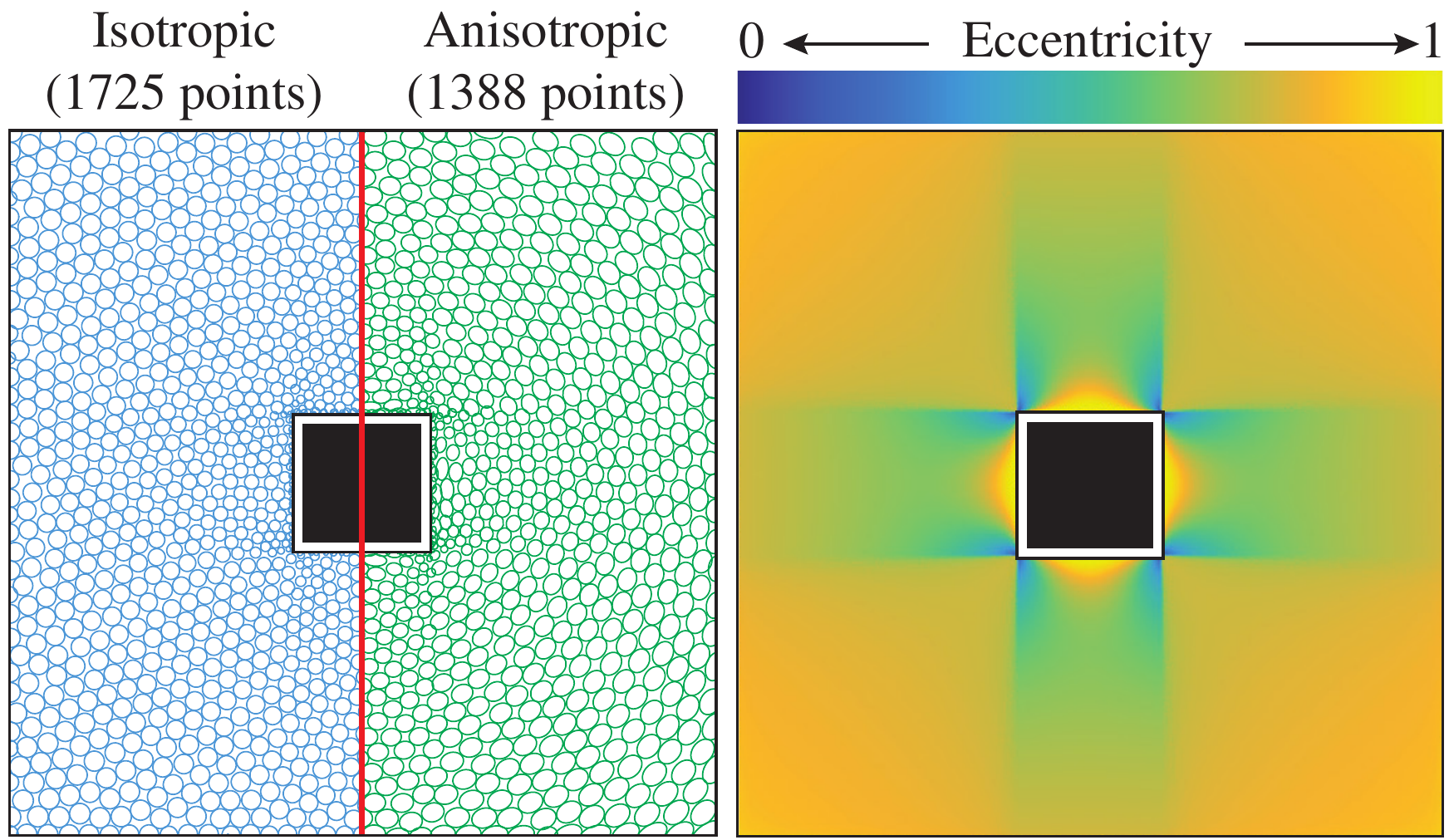}
	\caption{2D point distributions \minor{in a medium illuminated by a square-shaped light}, using our Hessian-based error metric with isotropic and anisotropic points (left) using the same relative error threshold. Eccentricity of radiance curvature
		(right) determines anisotropy of cache points (left, green), stretching circular points to ellipses along the direction of lower change.}
	\label{fig:isoAniso}
\end{figure}

By computing principal components of radiance Hessians, we can use the \emph{radiance eccentricity} (i.e. the eccentricity of the ellipse defined by the Hessian of the radiance) to stretch media cache points along the components with lower radiance variation, obtaining elliptic (2D) or ellipsoidal (3D) cache points. In \Fig{crossShadows} (bottom) we compare previous work with our isotropic and anisotropic cache distributions. Even with a similar number of isotropic points ($\sim$13k), our improved derivatives manage to capture the overlapping shadows much better; using our anisotropic technique, we manage to reduce cache size by 32\%, while keeping the same error threshold. \Fig{isoAniso} illustrates eccentricity across a 2D scene with a square light emitter in the center. By keeping the same error threshold, our anisotropic cache reduces the number of cache points by up to 20\%.

\subsection{Results in 3D}
Here we further analyze occlusion-unaware gradients and our oc\-clusion-aware Hessians on four 3D scenes: \emph{Strips}, \emph{Statues}, \emph{Patio} and \emph{Cornell holes}. Unless stated otherwise, all renders are taken using 16 samples per pixel, and performing uniform ray marching with a step size of 0.1.

The \emph{Statues} scene shown in \Fig{statues} combines both surface-to-media single scattering, and media-to-media (two-bounce) multiple scattering. The scene includes distant and local light sources (side windows and ceiling, respectively). Occlusion-unaware single and multiple scattering gradients lead to big splotches on the boundaries of light beams coming through the windows. In the case of light coming through the ceiling, while the point distribution captures shadow contours fairly well, extrapolation fails since occlusion-unaware gradients ignore light effects produced in the penumbra region. Moreover, occlusion-unaware techniques concentrate most cache points near light sources and reflecting surfaces (\Fig{statues}, middle), as seen previously in 2D. Since the gradients are large in these areas, this results in very small valid radii for the cache points. Histograms (\Fig{statues}, right) show how for previous work nearly 8000 points (leftmost bin, top blue histogram) on single scattering reach the minimum radius, which is close to a 40\% of the total number of points. This implies that the performance of this approach is highly dependent on the value of such minimum radius, which undermines the principled basis of its error metric. In contrast our method generates better point distributions, which correctly capture light gradients while avoiding additional heuristics to control oversampling in certain regions.

\begin{figure}[t]
	\centering
	\includegraphics[width=\columnwidth]{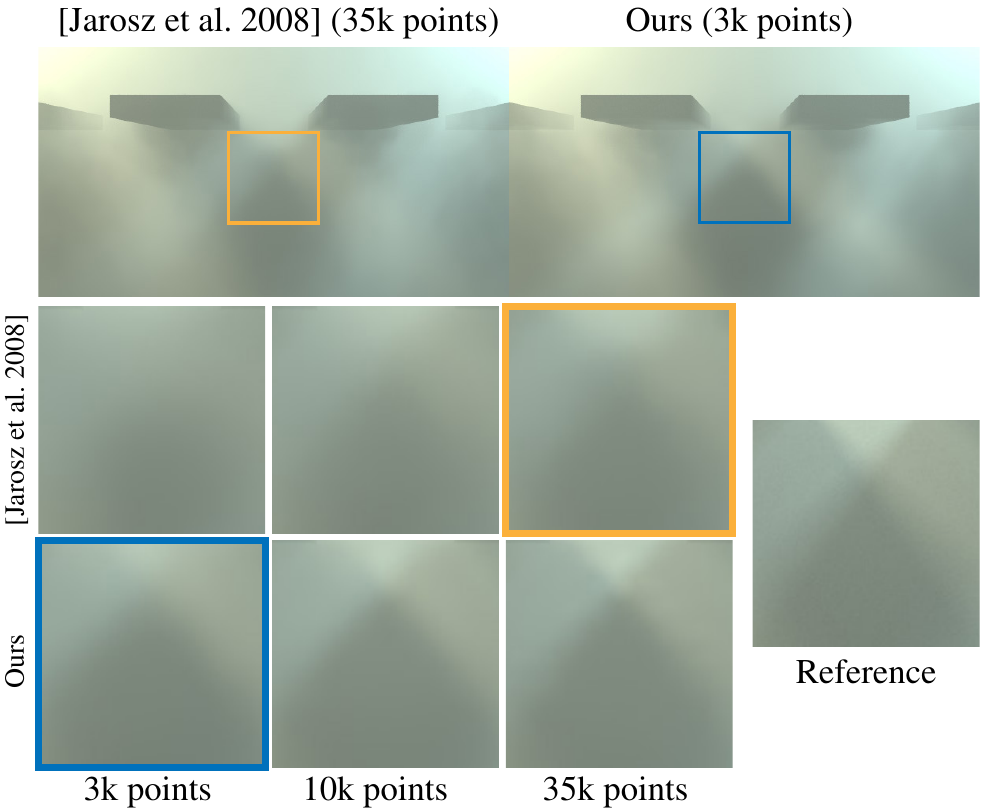}
	\caption{\textit{Strips} scene comparing single scattering for an increasing number of cache points, and showing relative error with respect to the reference image. While occlusion-unaware gradients method requires 35k cache points to fairly capture occlusions, our occlusion-aware Hessians produce similar results with just 3k points.}
	\label{fig:strips}
\end{figure}

The \emph{Strips} scene (\Fig{strips}) shows surface-to-medium single scattering, for an increasing number of cache sizes. Surface radiance is excluded for illustration purposes. The occlusion-unaware method needs an order of magnitude more cache points to get comparable results to ours (see progression insets). 
This implies that we have to significantly drop the tolerance parameter to create sufficiently fine point distributions in occluded regions.
%
%
As we can observe in \Fig{strips}, top row, our method yields better sampling density and extrapolation from the sampled points, achieving similar results with an order of magnitude less points.

\begin{figure}[t]
	\centering
	\includegraphics[width=\columnwidth]{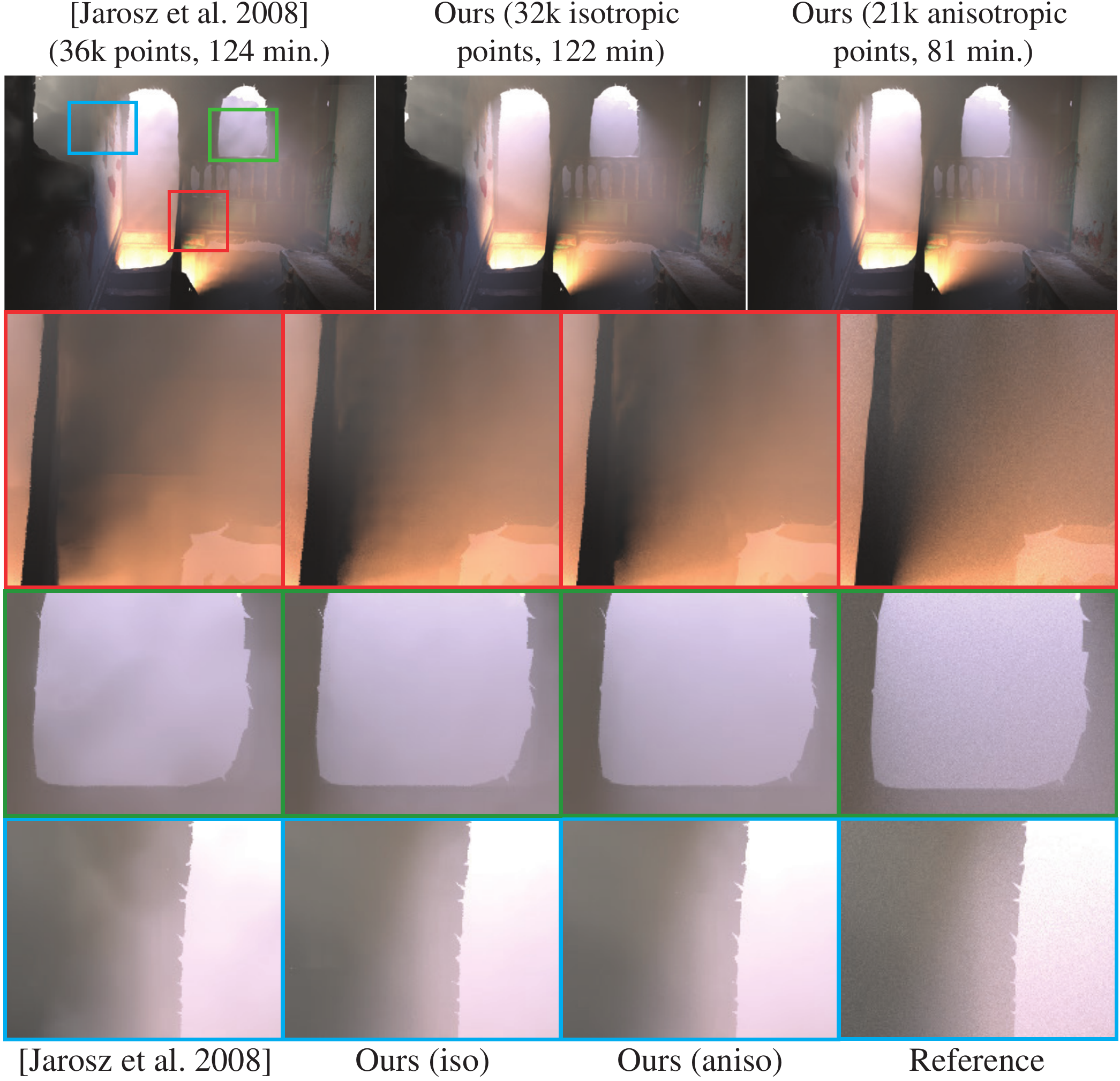}
	\caption{\emph{Patio} scene with single scattering. Our method outperforms existing occlusion-unaware techniques on an equal-time comparison. Moreover, our anisotropic cache manages to significantly reduce total time under the same error tolerance $\sErrorTolerance\! =\! 1.5\mathrm{e}{-4}$ than our isotropic cache, while still retaining shadow details on window boundaries and near thin handrails as shown in the insets.}
	\label{fig:patio}
\end{figure}

Computing derivatives of surface-to-medium form factor involves operating with $3\!\times\!3$ matrices (see \Apx{fluence3Dsegment}). Including the cost of scene subdivision, this introduces an overhead per cache point of just 9\%, compared to computing only point-to-point first derivatives (see \Tab{patioStats} for the \emph{Patio} scene).
Nevertheless, as we can see in \Fig{patio}, our method yields better equal-time results with isotropic points. Moreover, our anisotropic approach stretching spherical cache points along the principal components of radiance, allows to reduce both the number of points and the total computation time by 30\% for the same error tolerance.

\interfootnotelinepenalty=10000
\begin{table}[t]
	\caption{Computation data for the \emph{Patio} scene. For the isotropic case, our method yields better results in equal time. Using anisotropic points provides a further 30\% computation time reduction at the same low error threshold due to the improved point distributions and larger valid regions. \adrianc{Are we having this table for all the scenes? Or just for the \emph{Patio}? Do we want it for all the complex ones (i.e. \emph{Statues, Patio, Cornell Holes, Stairs, Whiteroom}?}}
	\label{tab:patioStats}
	\centering
	{
		\resizebox{\columnwidth}{!}{%
			\begin{tabular}{@{\;}l@{\;\;\;}c@{\;\;\;}r@{\;}c@{\;\;}r@{\;}}
				\toprule
				\textbf{Method} & \textbf{Error tol.\footnote{Note that Jarosz et al.'s metric (\Eq{jaroszRadius}) is different from our Hessian-based integrated error $\sErrorTolerance$, thus tolerance values of both metrics have different meaning.} \remove{\footnote{Jarosz and colleagues define an error metric based on the log-space gradient of radiance $\GRx L/ L$, different from our Hessian-based integrated error $\sErrorTolerance$, thus tolerance values of both metrics have different meaning. Please refer to \cite[Section 5.2]{jarosz08radiance} for more details.}}}  & \multicolumn{1}{c}{\textbf{Cache gen.}} & \textbf{Time / point} & \textbf{Total time} \\
				\midrule
				Jarosz et al.~2008 & $0.3$ &  124 min / 36k pts & 206 ms & 136 min \\
				Ours (isotropic) & $\sErrorTolerance\! =\! 1.5\mathrm{e}{-4}$ & 122 min / 32k pts & 225 ms  &  135 min\\
				Ours (anisotropic) & $\sErrorTolerance\! =\! 1.5\mathrm{e}{-4}$ &  81 min / 21k pts& 225 ms & 94 min \\
				\bottomrule
			\end{tabular}%
		}
	}
\end{table}

\begin{figure}[t]
	\centering
	\includegraphics[width=\columnwidth]{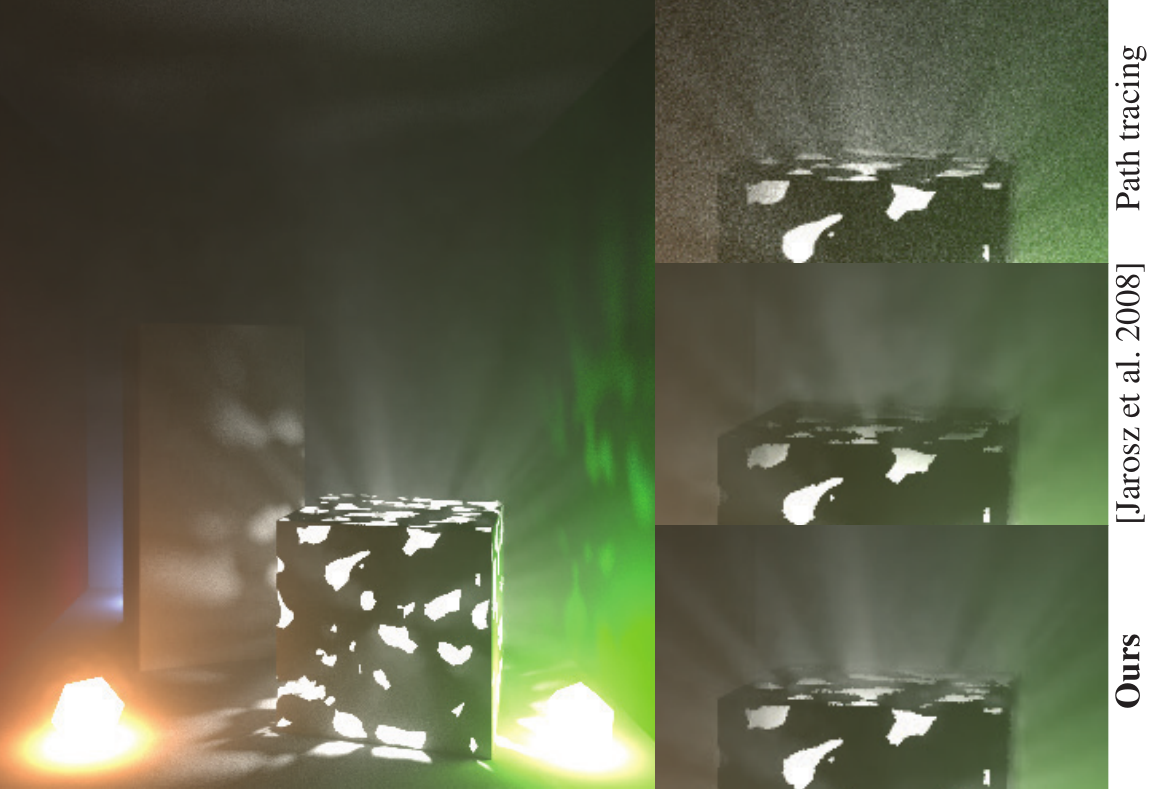}
	\caption{\textit{Cornell Holes} scene showing complex high-frequency shadows handled by our method, compared against equal-time path traced, and occlusion-unaware gradients solutions.}
	\label{fig:cornellHoles}
\end{figure}

The \emph{Cornell Holes} scene (\Fig{cornellHoles}) shows how our method successfully resolves difficult, high-frequency occlusions due to light coming out of the box. Our method provides a built-in mechanism to significantly reduce error in two ways: additional samples reduce variance but also create finer subdivisions, thus improving accuracy when detecting occlusions.

%

\begin{figure*}[t]
	\centering
	\includegraphics[width=\textwidth]{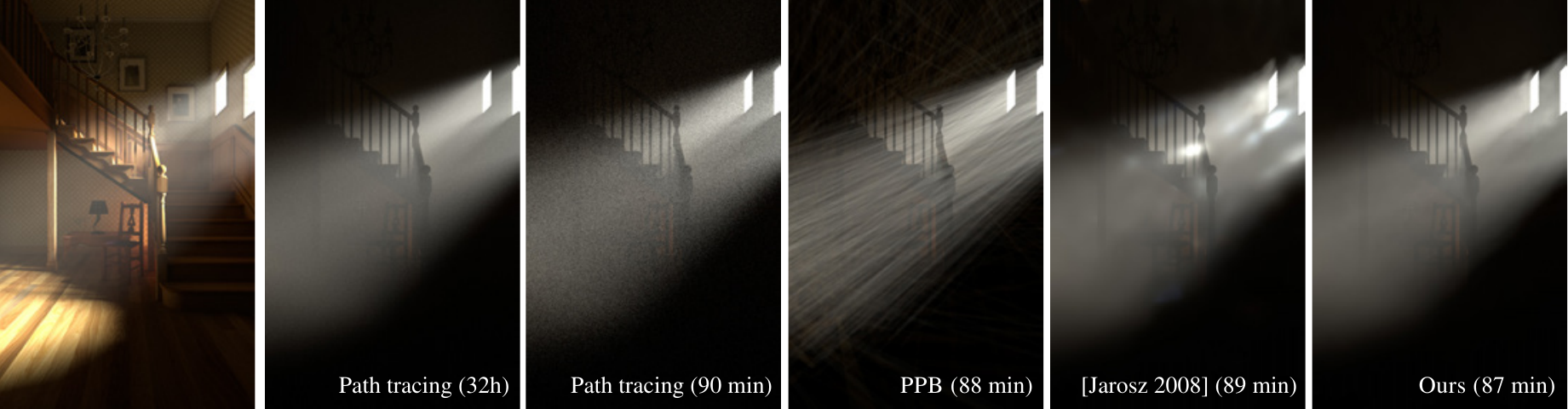}
	\caption{\major{\textit{Staircase} scene showing equal-time comparisons of path tracing, \minor{progressive photon beams (PPB)}, occlusion-unaware gradients~\protect{\citep{jarosz08radiance}}, and our second-order occlusion-aware solution. We include a fully converged solution for path tracing. \minor{Each cache in both our method and Jarosz et al. is computed using 16k stratified angular samples, and rendered using 16 samples per pixel. The progressive photon beams solution was obtained using the publicly available Tungsten rendering engine~\protect{\citep{bitterli2016resources}}}. Note how the occlusion unaware method creates visible artifacts in the patterns created by the shadows crossing from different windows, while our method correctly captures those details in equal time. \minor{Due to distant lighting, progressive photon beams fails to densely sample the light shafts coming through the windows, resulting in visible variance after 400 iterations of 1M beams/iteration.}}} 
	\label{fig:staircase}
\end{figure*}

\begin{figure}[t]
	\centering
	\includegraphics[width=\columnwidth]{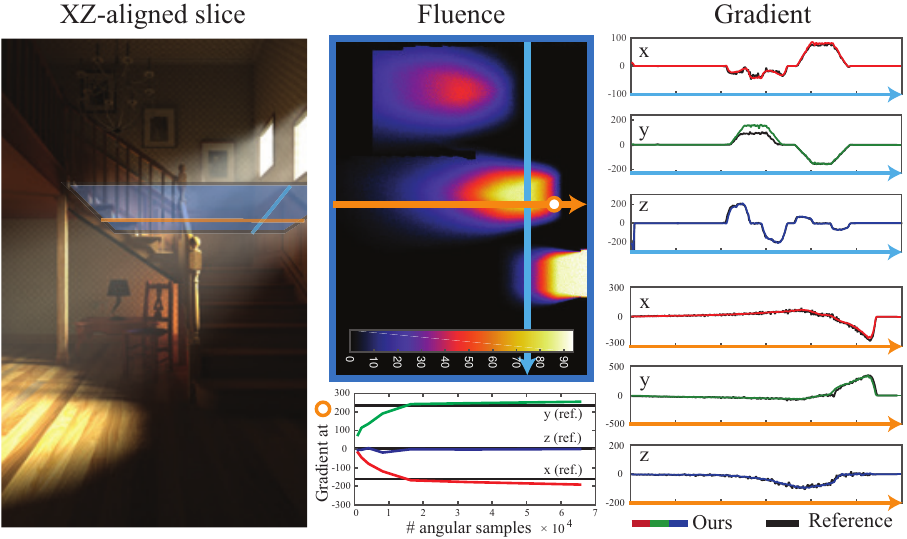}
	\caption{\major{We demonstrate the convergence of our occlusion aware derivatives in complex 3D scenarios like \textit{Staircase}. We illustrate this using an XZ-aligned slice of the media that captures the occlusion changes produced by the light shafts through the windows. Right graphs show our computed gradients across two orthogonal scan lines of the slice, where we can observe how our method matches the reference derivatives computed with finite differences. In the bottom graph we also illustrate convergence at the white dot respect to the number of angular samples. Higher number of angular samples create finer scene subdivisions and increase the precision of our derivatives, which provide a very good estimation of the actual derivatives.}}
	\label{fig:staircase_derivatives}
\end{figure}

\begin{figure*}[t]
	\centering
	\includegraphics[width=\textwidth]{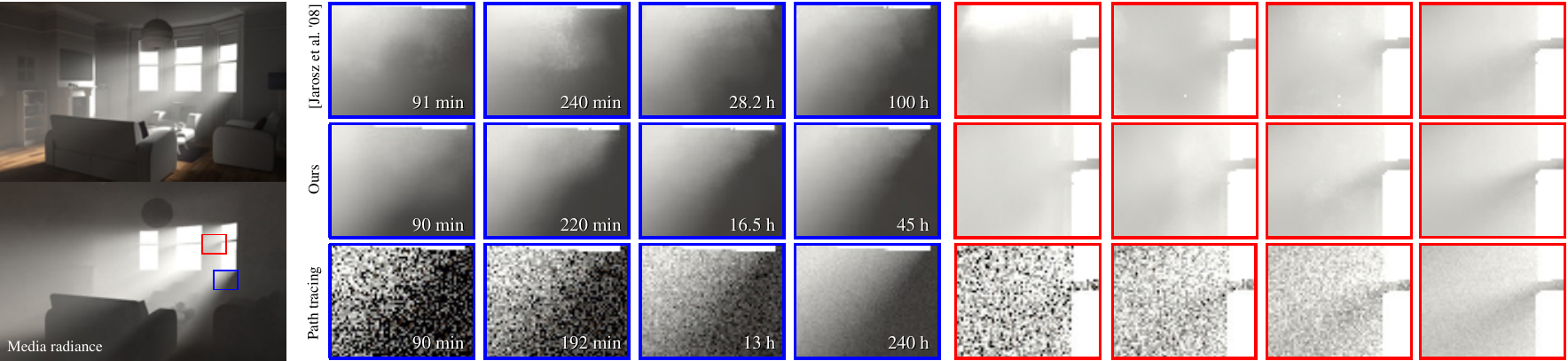}
	\caption{\major{We illustrate convergence to an equal-quality reference solution with our algorithm, Jarosz et al. method, and path tracing for the \textit{Whiteroom} scene. In 3D scenes of higher complexity, our method presents much better convergence properties than previous ones, being able to reconstruct the shadow boundaries near the window frames in much less time.}}
	\label{fig:whiteroom}
\end{figure*}

\major{We also demonstrate the benefits our method in scenes of higher complexity. In the \emph{Staircase} scene (\Fig{staircase}) we show an equal-time comparison with a render time of 90 minutes. Path tracing has not fully converged to the reference solution in that time, and while the point distributions of occlusion-unaware methods manage to capture the main shadow boundaries, occlusion-unaware gradients still create visible artifacts on the shadow patterns created by light coming from different windows. \minor{Progressive photon beams~\protect{\citep{jarosz11progressive}} manages to capture high frequency changes, but fails to densely sample the medium due to distant lighting}. In equal time, our method manages to get the closest match to the reference by correctly capturing complex shadow configurations. In \Fig{staircase_derivatives} we also illustrate convergence of our occlusion-aware gradients in the same scene by analyzing the changes on a XZ-aligned slice of the media crossing through the light shafts. We compare our gradients against finite differences gradients on two orthogonal scanlines that cross through the shadows, and demonstrate how our method converges to the reference gradients by creating finer subdivisions with higher number of angular samples. 

Finally we perform comparisons up to equal-quality in the \emph{Whiteroom} scene (\Fig{whiteroom}), which presents high scattering due to bright white walls and furniture. In a sequence of insets with increasing render time, we show how our method manages to recover high-frequency shadows in much less time than other methods, which also fail to capture thin shadows near window boundaries. }

\section{Conclusions}
\label{sec:discussion}
We have presented a new occlusion-aware method for efficiently computing light transport in homogeneous isotropic media, including both single and multiple scattering.
%
At the core of our method lies an efficient computation of radiance derivatives for both surface-to-medium and medium-to-medium light transport.
Our radiance derivatives, including visibility changes for single and multiple scattering, improve both the placement of cache points, as well as their interpolation using a Taylor expansion.

We have additionally formalized light transport in participating media in a self-contained 2D world; we hope that this  framework becomes a valuable contribution for the graphics community as a testbed for novel algorithms. Our results (2D and 3D) demonstrate a significant improvement over the current state of the art, both in equal-time and equal-error comparisons. \diegoc{check before submission}

\paragraph*{Limitations \& Future Work}
Our work shares some of the limitations of traditional radiance caching algorithms, namely the assumption of relatively low frequency transport \major{with finite derivatives. High-frequency illumination due to e.g.\ small light sources would require a very fine-grained subdivision to accurately find shadow boundaries. Other high-frequency effects such as caustics would additionally require departing from the assumption of constant angular radiance $\Rad_o$ in \Eq{ApproxSegmentRadiance}, which would in turn require computing its translational derivatives.}

\major{
In our implementation we have assumed isotropic media, which helps reduce the complexity and storage requirements of the cache points. By using an angularly-resolved caching of radiance and its derivatives (by using e.g.\ spherical harmonics~\citep{Krivanek2005caching,jarosz08radiance}) anisotropic phase functions could be added. }
\major{Incorporating heterogeneous media would break the assumption of constant scattering term (i.e. $\GRx\pf\neq\HSx\pf\neq0$) given the variability of $\sca$ and $\pfs$ within the media. This would require us to use the full radiance derivatives (\Eqs{SSSegALGradient}{SSSegALHessian}), instead of the simplified \Eqs{SSSegALGradientSimp}{SSSegALHessianSimp}. Moreover, it would require changing our derivatives of transmittance $\Tr$; given our marching procedure for subdividing the media, a similar approach to \citeauthor{jarosz08radiance}'s~\citeyear{jarosz08radiance} for single scattering could be used. Finally, high-frequency heterogeneity in the medium would require a very fine subdivision, which would potentially make our approach impractical.}

Our error metric assumes that the error is due to extrapolation only, with perfect radiance samples and derivatives. However, both are computed stochastically, which introduces variance (in the case of radiance), and bias (on the derivatives). Developing new metrics taking into account these additional sources of error, as well as accurately characterizing them, are interesting avenues of future work. In this regard, analyzing other consistent approaches to compute derivatives (e.g. using photon mapping~\citep{Kaplanyan2013APPM}) might be helpful. \major{Evaluating whether using our biased estimator of radiance (\Eq{ApprxRadiance}) instead of our Monte Carlo estimate of \Eq{outLF} would be interesting too, making our cache points more robust by reducing variance (at the price of additional bias). } \major{Finally, it may be possible to use our first- and second-order derivatives to accurately estimate the optimal kernel in density estimation algorithms for participating media~\citep{hachisuka13starpm}, as well as to guide sampling in media or to improve quadrature-based ray-marching methods \citep{Munoz14cgf}.}



\section*{Acknowledgements}
We want to thank the reviewers for their insightful comments. We also want to thank M. S. Suraj for discussions and help during early stages of the project, Ib\'{o}n Guill\'{e}n for helping with figures, and Benedikt Bitterli for his help on the comparisons with progressive photon beams. This project has received funding from the European Research Council (ERC) under the European Union's Horizon 2020 research and innovation programme (CHAMELEON project, grant agreement No 682080), DARPA (project REVEAL), and the Spanish Ministerio de Econom\'{i}a y Competitividad (projects \mbox{TIN2016-78753-P} and {TIN2014-61696-EXP}). Julio Marco was additionally funded by a grant from the Gobierno de Arag\'{o}n.

\bibliographystyle{ACM-Reference-Format}
\bibliography{paper-bib}

\appendix 
\section*{Appendices}
In the following we summarize 2D and 3D expressions of translational derivatives of transmittance and form factors needed for our method. 
We box all relevant final expressions that to the best of our knowledge are new to the literature. 
We define column vectors as $\vvec$ and row vectors as $\vvec^\mytransp$.  
Expressions such as  $\dotp{\rvec_1}{\rvec_2}$ denote dot (inner) products, while expressions such as  $\rvec_1 \rvec_2^{\;\mytransp}$, $\GRx(\ldots) \GRxT(\ldots)$, and $(\ldots)(\ldots)^\mytransp$ denote vector \emph{outer} products. 
%
\section{Homogeneous transmittance derivatives}
\label{ap:transmittance}
Homogeneous transmittance is modeled by the exponential decay due to extinction,
\begin{align}
T_r & = e^{-\ext \normyx}
\end{align}
where $\norm{\xy}$ denotes distance between source $\y$ and shaded point $\xm$. Its gradient and Hessian with respect to a translation of $\xm$ are
\begin{align}
\GRx T_r &=-\ext \fr{\yx}{r} T_r,
\end{align}%
\balign{%
\HSx T_r
&= -\ext (\fr{\JCx(\yx)}{r} - \fr{1}{r^3} \yx\,\yxT -  \fr{\ext}{r^2} \yx\,\yxT)\; \Tr.
}

\section{2D segment-media form factor derivatives}
\label{ap:segMediaFormFactor2D}
The form factor between a 2D segment $\ell$ and a media point $\xm$ (Figure~\ref{fig:setup2D3D}, left) is defined as the integrated curve-media geometry term along all segment points. This is equivalent to the angular ratio covered by $\seg$ as seen from~$\xm$
\begin{align*}
\FFDDM (\x)
&= \frac{1}{2\pi}\int_{\y_0}^{\y_1} \fr{\cosy}{\norm{\x-\y}}{\diffseg(\y)} = \frac{1}{2\pi}\arccos \left(\frac{\xy_0}{r_0} \cdot \frac{\xy_1}{r_1}\right).
\end{align*}
where $r_i = \norm{\xy_i}$.
The form factor gradient and Hessian become
\newcommand{\FFDDMThLims}[1]{#1^{\theta_\x^+}_{\theta_\x^-}}
\newcommand{\FFDDMYLims}[1]{#1^{\y_1}_{\y_0}}
\balign{
\GRx \FFDDM (\x)
&= -\frac{1}{2\pi}\frac{\GRx \costhetap}{\sqrt{1-\cos^2{\thetap}}}
\label{eq:SegToMediaFFGradient}
}
\balign{
\HSx \FFDDM (\x)
&=-\frac{1}{2\pi}\left(\frac{\JCx (\GRx \costhetap)}{\sqrt{1-\cos^2{\thetap}}}\right. \nonumber \\
&+\left.\frac{\costhetap}{(1-\cos^2{\thetap})^{3/2}} \GRx \costhetap \GRxT \costhetap\right)
\label{eq:SegToMediaFFHessian}
}
where \major{$\JCx$ is the Jacobian operator}, and:
\begin{align}
\GRx \costhetap
&= \frac{\costhetap}{r_0^2}\xy_0+\frac{\costhetap}{r_1^2}\xy_1\\
&- \frac{(\xy_0 + \xy_1)}{r_0 r_1},\\
%
\JCx(\GRx \costhetap)
&= -\JCx\left(\frac{\xy_0}{r_0 r_1}\right) - \JCx\left(\frac{\xy_1}{r_0 r_1}\right) \nonumber \\
& + \JCx\left(\frac{\costhetap}{r_0^2} \xy_0 \right) + \JCx\left(\frac{\costhetap}{r_1^2} \xy_1 \right),\\
%
\JCx\left(\frac{\xy_i}{r_0 r_1}\right) &= \frac{\JCx(\xy_i)}{r_0 r_1}+ \frac{\xy_i \xyT_0 }{r_0^3 r_1} + \frac{\xy_i \xyT_1}{r_0 r_1^3},\\
%
\JCx \left(\frac{\costhetap}{r_i^2} \xy_i \right)
&= \frac{\costhetap}{r_i^2} \JCx(\xy_i) + \frac{\xy_i}{r_i^2} \GRxT \costhetap \nonumber \\
&+ \frac{2 \costhetap}{r_i^4}\xy_i \xyT_i.
\end{align}


\newcommand{\ririi}{\vv{\mathbf{\varrho}}_{\! i}}
\newcommand{\RN}[1]{%
	\textup{\uppercase\expandafter{\romannumeral#1}}%
}
\newcommand{\mtxform}[1]{\langle #1 \rangle}

\section{3D triangle-media form factor derivatives}
\label{ap:fluence3Dsegment}
\renewcommand{\vvec}{\vv{\mathrm{\textbf{v}}}}
\begin{figure}[t]
	\centering
	\def\svgwidth{0.7\columnwidth}
	\small
	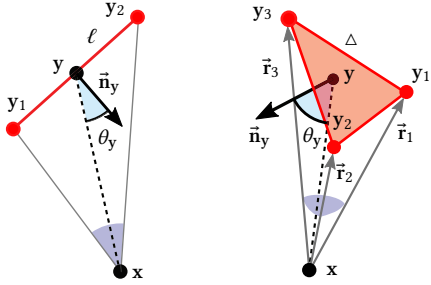
	\caption{Setups for segment-to-media (2D, left) and triangle-to-media (3D, right) form factors.}
	\label{fig:setup2D3D}
\end{figure}
The form factor between a 3D triangular face $\tri$ and a media point $\x$ (see Figure~\ref{fig:setup2D3D}, right) is defined as the integrated surface-media geometry term along all points in the triangle. Analogous to 2D, this has analytical solution equal to the ratio of solid angle covered by the triangle as seen from $\x$,
\begin{align}
\FFDDDM (\x)
&= \frac{1}{4\pi}\int \limits_{\y \in \tri} \fr{\cosy}{\norm{\x-\y}^2}{\diff \tri(\y)} = \frac{\SATriangle}{4\pi}.
\label{eq:FFDDDMediaSurfArea}
\end{align}
Solid angle $\SATriangle$ of a triangle can be computed as \cite{Oosterom1983},
\newcommand{\absA}{\absvalue{A}}
\begin{align}
\SATriangle
&= 2\arctan{\frac{\absA}{B}}\\
\intertext{with}
A & = {\dotp{\rvec_1 }{\left(\rvec_2\times\rvec_3\right)}} \\
B & = {\rlen_1 \rlen_2 \rlen_3 + \dotpp{\rvec_1}{\rvec_2}\rlen_3 + \dotpp{\rvec_2}{\rvec_3}\rlen_1 + \dotpp{\rvec_1}{\rvec_3}\rlen_2}
\end{align}
where $\rvec_i = \xyi$, and $\rlen_i = \norm{\rvec_i}$ (see Figure~\ref{fig:setup2D3D}, right). Note that the numerator $A$ requires an absolute value to ensure positive vector order (i.e.\ triangle winding) with respect to $\xm$. Also, when obtaining negative arctangent values, $\pi$ must be added to the obtained solid angle.

The gradient of the form factor with respect to a translation of \x becomes
\balign{
\GRx \FFDDDM (\x)
&=\frac{1}{2\pi} \GRx \arctan{\frac{\absA}{B}}\\
&=\frac{1}{2\pi} \frac{B \GRx\absA - \absA \GRx B}{\absA^2 + B^2},
\label{eq:TriToMediaFFGradient}
}
and its Hessian yields
\balign{
	\HSx \FFDDDM (\x)& =\frac{1}{2\pi} \left( \frac{\GRx(\absA) \GRxT B - \GRx B \GRxT (\absA)}{\absA^2 + B^2} \right. \eqbreak
	& +\! \frac{B\JCx(\GRx (\absA)) - \absA \JCx (\GRx B)}{\absA^2 + B^2 }  \eqbreak
	& -\! \left.\frac{\left(B \GRx(\absA)\! -\! \absA \GRx B\right) \left(\GRx(\absA^2) + \GRx (B^2)\right)^\mytransp}{\left(\absA^2 + B^2\right)^2}\! \right)\!\!.\!
	\label{eq:TriToMediaFFHessian}
}
Note that for computing the terms $\GRx (\absA)$ and $\JCx(\GRx \absA)$, we can apply the derivatives of the absolute value of a vector function:
\begin{align}
&\GRx (\absA) = \frac{A}{\absA} \GRx A, \\
&\JCx(\GRx \absA) = \frac{A\JCx(\GRx A) + \GRx A \GRxT A}{\absA} - \frac{A^2 (\GRx A \GRxT A)}{\absA^3}.
\end{align}

The gradient of $A$ becomes
\begin{align}
\GRx A
&= {\JCx\left(\rvec_2\times\rvec_3\right)}{\rvec_1} + \JCx\left(\rvec_1\right)\crosspp{\rvec_2}{\rvec_3}.
\end{align}
By the Jacobi identity we have that
\begin{align}
\JCx \crosspp{\rvec_2}{\rvec_3}
&= \rvec_2 \times \JCx (\rvec_3)  - \rvec_{3} \times \JCx (\rvec_2)
\label{eq:JCr1r2}
\end{align}
where any vector-matrix cross product $\vvec \times \JCx(\bullet)$ can be expressed by means of the matrix multiplication form
\begin{align}
\label{eq:vectormatrixcross}
&\vvec \times \JCx(\bullet) = \mtxform{\vvec}  \JCx(\bullet) \\
\label{eq:vmtxform}
&\vvec = \mat{v_\textrm{(1)}\\ v_\textrm{(2)}\\ v_\textrm{(3)}}, \qquad  \mtxform{\vvec}  = \mat{0 &-v_\textrm{(3)} &v_\textrm{(2)} \\ v_\textrm{(3)} &0 &-v_\textrm{(1)} \\ -v_\textrm{(2)} & v_\textrm{(1)} & 0}.
\end{align}
Since $\JCx(\rvec_1)\! =\! \JCx(\rvec_2)\! =\! \JCx(\rvec_3)\! =\! -I_{3}$, we have that
\balign{
\GRx A
&= \left(\mtxform{\rvec_2} \JCx (\rvec_3)-\mtxform{\rvec_3} \JCx (\rvec_2)\right) {\rvec_1} - \crosspp{\rvec_2}{\rvec_3}\eqbreak
&= \mtxform{\rvec_3 - \rvec_2}{\rvec_1} - \crosspp{\rvec_2}{\rvec_3}.
}
Note that $\mtxform{\rvec_3 - \rvec_2} = \mtxform{\y_3 - \y_2}$ and therefore does not depend on $\xm$, and $\mtxform{\vvec}^\mytransp = \mtxform{-\vvec}$ (see \Eq{vmtxform}). As a result, the Jacobian of $\GRx A$ becomes a zero matrix
\balign{
\JCx \left(\GRx A \right)
&= \JCx({\rvec_1})\mtxform{\rvec_3 - \rvec_2} ^\mytransp - \JCx\crosspp{\rvec_2}{\rvec_3} \eqbreak
&= \mtxform{\rvec_3 - \rvec_2} - \mtxform{\rvec_3 - \rvec_2} \eqbreak
&= 0.
}

The gradient of $B$ becomes
\balign{
\GRx B
&= \GRx(\rlen_1 \rlen_2 \rlen_3) + \GRx\left(\dotpp{\rvec_1}{\rvec_2}\rlen_3\right) \eqbreak
&+ \GRx\left(\dotpp{\rvec_2}{\rvec_3}\rlen_1\right) +\GRx\left(\dotpp{\rvec_1}{\rvec_3}\rlen_2\right)
}
where
\begin{align}
&\GRx(\rlen_1 \rlen_2 \rlen_3)
= \rlen_2 \rlen_3 \GRx \rlen_1 +\rlen_1 \rlen_3 \GRx \rlen_2+\rlen_1 \rlen_2 \GRx \rlen_3 \\
&\GRx \left(\dotpp{\rvec_i}{\rvec_j}\rlen_k\right) = \dotpp{\rvec_i }{\rvec_j} \GRx \rlen_k - \rlen_k (\rvec_i + \rvec_j)\\
&\GRx \rlen = -\frac{\rvec}{\rlen} .
\end{align}

Jacobian of $\GRx B$ yields
\balign{\JCx (\GRx B)
&= \JCx (\GRx(\rlen_1 \rlen_2 \rlen_3)) + \JCx\left(\GRx\left(\dotpp{\rvec_1}{\rvec_2}\rlen_3\right)\right) \eqbreak
&+ \JCx\left(\GRx\left(\dotpp{\rvec_2}{\rvec_3}\rlen_1\right)\right) +  \JCx\left(\GRx\left(\dotpp{\rvec_1}{\rvec_3}\rlen_2\right)\right)
}
where
\begin{align}
\JCx (\GRx(\rlen_1 \rlen_2 \rlen_3))
& =\rlen_2\rlen_3\JCx(\GRx \rlen_1)  +\rlen_1(\GRx \rlen_3 \GRxT \rlen_2 + \GRx \rlen_2 \GRxT \rlen_3) \eqbreak
& +\rlen_1\rlen_3 \JCx(\GRx \rlen_2) + \rlen_2 (\GRx \rlen_3 \GRxT \rlen_1 + \GRx \rlen_1 \GRxT \rlen_3)  \eqbreak
& + \rlen_1 \rlen_2 \JCx(\GRx \rlen_3) + \rlen_3(\GRx \rlen_2 \GRxT \rlen_1 + \GRx \rlen_1 \GRxT \rlen_2)
\end{align}
\begin{align}
\JCx\left(\GRx\left(\dotpp{\rvec_i}{\rvec_j}\rlen_k\right)\right)
&=  \dotpp{\rvec_i}{\rvec_j}\JCx(\GRx \rlen_k)�+2 \rlen_k I_3 \eqbreak
&-\GRx \rlen_k  (\rvec_i + \rvec_j)^\mytransp - (\rvec_i + \rvec_j) \GRxT \rlen_k
\\
\eqbreak
\JCx(\GRx \rlen) &= \frac{I_3}{\rlen} -\frac{\rvec\; \rvec^\mytransp}{\rlen^3}.
\end{align}



\section*{Supplementary material}

# Second-Order Occlusion-Aware Volumetric Radiance Caching: Supplementary Material

JULIO MARCO, Universidad de Zaragoza, I3A
ADRIAN JARABO, Universidad de Zaragoza, I3A
WOJCIECH JAROSZ, Dartmouth College
DIEGO GUTIERREZ, Universidad de Zaragoza, I3A

This is the supplementary material for the paper *Second-Order Occlusion-Aware Volumetric Radiance Caching*. The document contains the following sections:

- **1 Geometry terms and their derivatives for 2D curve-medium and 3D surface-medium.**
- **2 Derivation of our second order metric for media radiance extrapolation in 2D and 3D**

## 1 GEOMETRY TERM DERIVATIVES

In order to compute Jarosz et al. [2008] translational gradients in 2D and 3D, we have to account for different curve-medium and surface-medium geometry terms respectively. As previously noted, in our defined self-contained world (Section 3), transmittance and its derivatives are equivalent in 2D and 3D. Please refer to Appendix A in the main text for their expressions. In the following we summarize Jarosz and colleagues's 2D and 3D geometry terms and their derivatives.

### 1.1 2D curve-medium

The geometry term between a curve (2D equivalent of surface) point $y_s$ and a 2D medium point $x$ is (see Figure 1, left)

$$G_{2D}(x, y_s) = \frac{\cos \theta_y}{r}$$
$$= \frac{\vec{yx} \cdot \vec{n}_y}{r^2}, \quad (S.1)$$

with $\vec{yx} = x - y_s$, $\vec{n}_y$ the normal at $y_s$, and $r = \|x - y_s\|$. Its translational gradient yields

$$\nabla G_{2D}(x, y) = \frac{\vec{n}_y}{r^2} - 2 \frac{\cos \theta_y}{r^3} \vec{yx}. \quad (S.2)$$

### 1.2 3D surface-medium

The geometry term between a surface point $y_s$ and a 3D medium point $x$ is (see Figure 1, right)

$$G_{3D}(x, y_s) = \frac{\cos \theta_y}{r^2}$$
$$= \frac{\vec{yx} \cdot \vec{n}_y}{r^3}, \quad (S.3)$$

while its translational gradient is

$$\nabla G_{3D}(x, y_s) = \frac{\vec{n}_y}{r^3} - \frac{3 \cos \theta_y}{r^4} \vec{yx}. \quad (S.4)$$

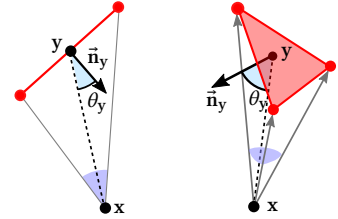

Fig. 1. Setups for curve-medium (2D, left) and triangle-medium (3D, right) geometry terms.

## 2 SECOND-ORDER ERROR CONTROL FOR MEDIA RADIANCE EXTRAPOLATION

The error control in radiance caching is mainly dependent on the error introduced by extrapolation from the radiance cache points. This error is controlled by an error tolerance value $\varepsilon$, and depends on how radiance is extrapolated in the cache neighborhood, as well as on the radiance moments at cache point $x$. These in the end give a valid bounding region $\aleph$ that determines whether a point $x'$ is suitable for extrapolation. Jarosz et al.'s [2008] existing radius computation technique is based on local contrast of media radiance, and only relies on first-order derivatives. In practice this leads to cache distributions of points that usually oversample radiance close to surfaces, while undersampling regions far from light sources (see main text, Figure 7).

Given the radiance and the first $n$ derivatives at a media point $x$, we can compute the approximated radiance at point $x' \in \aleph$ using a $n^{\text{th}}$-order Taylor expansion as:

$$\hat{L}_n(x') = \sum_{j=0}^{n} \frac{L^{(j)}(x)}{j!} (\Delta_{x'})^j \quad (S.5)$$

where $\Delta_{x'} = x' - x$, and $L^{(j)}(x)$ indicates the $j^{\text{th}}$ translational derivative of radiance at $x$. This approximation introduces an error

$$\epsilon_n(x') = |L(x') - \hat{L}_n(x')|, \quad (S.6)$$

which is the remainder of the Taylor expansion of order $n$. Unfortunately, we cannot compute the exact value of $\epsilon_n(x')$, since this would require knowing the exact value of the radiance $L(x')$. Moreover, given that we cannot guarantee continuity of $L^{(j)}$ in the closed interval between $x$ and $x'$, and its immediately higher-order derivative $L^{(j+1)}$ is unknown, we cannot compute explicitly this error using e.g. its Lagrange or its integral form.

20:2 • Marco et al.

Instead, we can use our best approximation of the actual radiance —the $n^{\text{th}}$-order Taylor expansion $\hat{L}_n(\mathbf{x}')$— to estimate the error of the $(n-1)^{\text{th}}$-order radiance approximation,

$$\epsilon_{n-1} \approx \hat{\epsilon}_{n-1}(\mathbf{x}') = |\hat{L}_n(\mathbf{x}') - \hat{L}_{n-1}(\mathbf{x}')|$$
$$= \left|\frac{L^{(n)}(\mathbf{x})}{n!}(\Delta_{\mathbf{x}'})^n\right|, \quad (S.7)$$

which results in an estimated error proportional to the highest known derivative at $\mathbf{x}$. This is equivalent to the Lagrange form of the remainder assuming that $L^{(n)}$ is constant in the open interval $(\mathbf{x}, \mathbf{x}')$. This assumption has been used in previous works truncating the Taylor expansion up to order $n = 1$ [Jarosz et al. 2008] and $n = 2$ [Jarosz et al. 2012; Schwarzhaupt et al. 2012], the latter for surfaces only.

Using this error estimation $\hat{\epsilon}_{n-1}$ we can compute the total (estimated) error introduced by a cache point as the integral over its valid region $\aleph$ as:

$$\hat{\epsilon}_{n-1}(\aleph) = \int_\aleph \hat{\epsilon}_{n-1}(\mathbf{x}')d\mathbf{x}'. \quad (S.8)$$

The introduced error is therefore proportional to the size of $\aleph$; by inverting this integral we can therefore compute the maximum region that satisfies $\hat{\epsilon}_{n-1}(\aleph) < \varepsilon$. To take into account the sensitivity to contrast of the human visual system, we aim to bound the *relative* error $\hat{\epsilon}'_{n-1}(\aleph) = \hat{\epsilon}_{n-1}(\aleph)/L(\mathbf{x})$.

*Truncation to Order 2 for Participating Media.* Existing work on radiance caching for participating media estimates relative error using radiance gradients at $\mathbf{x}'$. However, ignoring higher order derivatives creates suboptimal cache distributions that often oversample regions near surfaces and light sources. Moreover, this error formulation does not match the radiance extrapolation based on Taylor expansion of order 1. By computing second-order derivatives of media radiance we improve the estimated relative error as

$$\hat{\epsilon}'(\mathbf{x}') = \frac{\left|\Delta_{\mathbf{x}'}^\mathsf{T} \mathbf{H}L(\mathbf{x})\Delta_{\mathbf{x}'}\right|}{2L(\mathbf{x})}, \quad (S.9)$$

where $\mathbf{H}L$ denotes the radiance Hessian; note that we have removed the expansion order to simplify notation. This approach has been successfully used before to estimate error in *surface* irradiance [Jarosz et al. 2012; Schwarzhaupt et al. 2012].

For $k$-D participating media, we can compute the estimated relative error for a first-order expansion as the volume integral on the valid region $\aleph$ around $\mathbf{x} = \{x_1 \ldots dx_k\}$ as

$$\hat{\epsilon}'(\aleph) = \frac{1}{2}\int_{\aleph \in \mathbb{R}^k} \frac{\left|\Delta_{\mathbf{x}'}^\mathsf{T} \mathbf{H}L(\mathbf{x})\Delta_{\mathbf{x}'}\right|}{L(\mathbf{x})} d\mathbf{x} \quad (S.10)$$

where the Hessian $\mathbf{H}L$ is a $k \times k$ matrix. The $k$ eigenvalues $\lambda_1 \ldots \lambda_k$ of the Hessian allow us to modulate the radius of the valid region differently along the eigenvectors of radiance, to obtain an anisotropic valid region for each cache point, with an error upper bound

$$\hat{\epsilon}'(\aleph) \leq \int \cdots \int_{\aleph \in \mathbb{R}^k} \frac{\sum_{i=1}^k |\lambda_i| x_i^2}{L(\mathbf{x})} dx_1 \ldots dx_k \quad (S.11)$$

where $x_1 \ldots x_k$ lay in the orthonormal space defined the Hessian eigenvectors $\{\vec{v}_{\lambda_1} \ldots \vec{v}_{\lambda_k}\}$. Therefore, the valid region for a cache point in a $k$-dimensional space is an hyperellipsoid. By integrating the valid region and inverting the inequality we can compute the maximum space on the Hessian eigenspace, satisfying that the relative error is below the threshold $\varepsilon$, as a set of $k$ maximum distances $R^{\lambda_1} \ldots R^{\lambda_k}$. Similarly, we can define a single maximum distance $R$ by choosing only the maximum absolute eigenvalue $|\lambda_1|$, defining a $k$-D hypersphere instead of a $k$-D hyperellipse.

For $k = 2$, we can parameterize the valid region for a given error threshold $\varepsilon$ as the ellipse with radii $R_{\text{2D}}^{\lambda_i}$:

$$R_{\text{2D}}^{\lambda_i} = \sqrt[4]{\frac{4L(\mathbf{x})\varepsilon}{\pi|\lambda_i|}}. \quad (S.12)$$

This formula is analogous to the relative error metric presented by Schwarzhaupt and colleagues [2012], where radii are computed by projecting surface irradiance Hessian over the surface 2D subspace and obtaining its principal components.

Similarly, for a more practical three-dimensional world ($k = 3$), the valid region for a cache point becomes an ellipsoid in 3D, parameterizing its radii $R_{\text{3D}}^{\lambda_i}$ as a function of the relative error threshold $\varepsilon$ as

$$R_{\text{3D}}^{\lambda_i} = \sqrt[5]{\frac{15L(\mathbf{x})\varepsilon}{4\pi|\lambda_i|}}. \quad (S.13)$$

Note that this second-order error metric and its derived radius formulae assume perfect known values of radiance and its derivatives at $\mathbf{x}$. In practice, these are usually computed by Monte Carlo tracing radiance across the scene. This leads to other sources of error of different nature, such as variance inherent to Monte Carlo sampling, or bias due to limitations in the derivatives computation method. In Sections 4 and 6 we describe and illustrate in more detail these sources of error.

While the presented metric describes error introduced by extrapolation from a single cache point, it is common practice to first pre-populate the cache by tracing (a subset) the pixels of the image, and then perform a so-called *overture* pass that renders the final image. This allows the shading values to be *interpolated* (instead of *extrapolated*) from suitable cache points, increasing reconstruction quality. Due to the nature of the metric, only points that ensure a maximum error $\hat{\epsilon}'$ are used to interpolate radiance at $\mathbf{x}'$, and therefore the error of the weighted sum is equally upper-bounded by $\hat{\epsilon}'$.


\end{document}